\documentclass[11pt,fleqn]{article}
\usepackage{rotating,latexsym,amsmath,amssymb, color}
\usepackage{longtable}
\usepackage{caption}
\usepackage{a4wide}
\usepackage{graphicx}
\usepackage{float}
\usepackage{multirow}
\usepackage{booktabs}
\usepackage{mathrsfs}
\usepackage{bm}
\usepackage{hyperref}

\usepackage{cancel}
\usepackage{soul}
\begin{document}
\numberwithin{equation}{section}
\renewcommand{\thefootnote}{*}
\renewcommand{\figurename}{\bf Fig.}
\thispagestyle{empty}

\begin{center}
\textcolor{black}{{\Large \Large\bf On the Unit Teissier Distribution:\\ Properties, Estimation Procedures and Applications}\\}

\vspace{.5cm}
{\bf Zuber Akhter$^1$\footnote{Corresponding author e-mail address: akhterzuber022@gmail.com}, Mohamed A. Abdelaziz$^2$, M.Z. Anis$^3$ and Ahmed Z. Afify$^2$}\\

\vspace{0.3cm}
$^1$Department of Statistics\\
University of Delhi, Delhi-110 007, India\\

\vspace{0.20cm}
$^2$Department of Statistics, Mathematics \& Insurance\\
Benha University, Benha-13511, Egypt\\

\vspace{0.20cm}
$^3$SQC \& OR Unit\\
Indian Statistical Institute, Calcutta-700108, India
\end{center}

\begin{abstract}
\noindent 
The Teissier distribution, originally proposed by Teissier \cite{Teissier1934}, was designed to model mortality due to aging in domestic animals. More recently, Krishna et al. \cite{Krishna2022} introduced the Unit Teissier (UT) distribution on the interval $(0,1)$ through the transformation $X = e^{-Y}$, where $Y$ follows the Teissier distribution. In their work, the authors derived several fundamental properties of the UT distribution and investigated parameter estimation using maximum likelihood, least squares, weighted least squares, and Bayesian methods. Building upon this work, the present paper develops additional theoretical and inferential results for the UT distribution. In particular, closed-form expressions for single moments of order statistics and L-moments are obtained, and characterization results based on truncated moments are established. Furthermore, several alternative parameter estimation methods are considered, including maximum product of spacings, Cramér–von Mises, Anderson–Darling, right-tail Anderson–Darling, percentile, and L-moment estimation, while the estimation methods previously studied by Krishna et al. \cite{Krishna2022} are also included for comparison. Extensive simulation studies under various parameter settings and sample sizes are conducted to assess and compare the performance of the estimators. Finally, the 
exibility and practical utility of the UT distribution are demonstrated using a real dataset.
\end{abstract}

{\bf Keywords:} Unit Teissier distribution; L-moments; Order statistics; Characterization, Parameter estimation.

\vspace{0.25cm}
{\bf 2020 AMS Subject Classification:} 62G30, 62E10, 62F10. 

\section{Introduction}
In recent years, considerable effort has been devoted to developing new probability distributions in order to address the increasing complexity of data encountered in practice. Rather than proposing entirely new models, many studies have focused on extending existing \textcolor{black}{the } classical distributions, as such extensions often provide greater flexibility while retaining mathematical tractability. These extended models have found wide applicability in areas such as reliability and lifetime analysis, engineering, economics, finance, demography, actuarial science and medical research, where data frequently exhibit diverse structural patterns.

\vspace{0.25cm}
Within this broad area of research, probability distributions defined on the unit interval (0,1) are of particular importance, as they are well suited for modeling proportions, rates and other bounded observations. Among such models, the beta distribution is one of the most widely used for describing response variables that take values in the interval $(0,1)$. Its widespread use is largely due to its flexibility and its ability to provide a satisfactory fit in many practical situations, and it has therefore been studied extensively in both theoretical and applied statistics for over a century. Nadarajah and Kotz~\cite{Nadarajah2007} presented a comprehensive review of the properties of the beta distribution and its various extensions, including its relationships with other distributions and applications to economic and drought data sets. Further discussions on applications of the beta distribution can be found in Johnson et al.~\cite{Johnson1995} and Gupta and Nadarajah~\cite{GuptaNadarajah2004}, and the references therein.

\vspace{0.25cm}
Although the beta distribution has long been the standard choice for such data, its analytical complexity especially the absence of closed form expressions for certain key functions has motivated the investigation of alternative unit interval models. Therefore, several distributions supported on (0,1) have been proposed in the literature, aiming to achieve an appropriate balance between flexibility and analytical convenience.

\vspace{0.25cm}
Several flexible distributions for modeling unit interval data have been proposed in the literature. Nakamura et al. \cite{Nakamura2019} introduced a unit distribution based on the sinh–arcsinh transformation, which is capable of capturing varying levels of skewness and kurtosis, and implemented it within the GAMLSS framework with an application to football points-ratio data. Sarhan \cite{Sarhan2025} proposed the Unit-Chen distribution along with a quantile regression framework for unit-interval responses. Mazucheli et al. \cite{Mazucheli2019} developed the Unit-Gompertz distribution and demonstrated its usefulness using maximum flood level data of the Susquehanna River and tensile strength data of polyester fibers. More recently, Ormoz et al. \cite{Ormoz2026} extended the Unit-Gompertz model by proposing a three-parameter reparameterisation and studied its properties under lower k-record values, developing linear estimation and prediction methods with an application to COVID-19 data. In a comprehensive contribution, Afuecheta et al. \cite{Afuecheta2025} reviewed over one hundred unit continuous distributions, including classical models such as the beta and Kumaraswamy distributions, emphasizing their theoretical properties and practical applicability.

\vspace{0.25cm}
Recently, Krishna et al. \cite{Krishna2022} introduced the Unit Teissier (UT) distribution through the transformation $X = e^{-Y}$, where $Y$ follows the Teissier distribution. The probability density function (pdf) for a positive shape parameter $(\theta > 0)$ is given by
\begin{equation}\label{eq:1.1}
f(x;\theta) = \theta \left( x^{-\theta} - 1 \right) x^{-(1+\theta)}
\exp\!\left(1 - x^{-\theta}\right), \quad 0 < x < 1,
\end{equation}
\noindent
and the corresponding cumulative distribution function (cdf) takes the form
\begin{equation}\label{eq:1.2}
F(x;\theta) = x^{-\theta} \exp\!\left(1 - x^{-\theta}\right), \quad 0 < x < 1.
\end{equation}
Henceforth, we write $X\sim \text{UT}(\theta)$ if the pdf of $X$ can written as \eqref{eq:1.1}. Notice that
\begin{equation}\label{eq:1.3}
x{f(x;\theta)}={\theta(x^{-\theta}-1)} {F(x;\theta)}
\end{equation}
Krishna et al. \cite{Krishna2022} investigated several fundamental properties of the UT distribution, such as moments, entropy measures, and hazard rate behavior. A notable advantage of the UT distribution is the availability of closed form expressions for both its pdf and cdf, which greatly facilitates theoretical developments and practical implementation. Despite its simple structure, the model is flexible enough to accommodate a wide variety of hazard rate shapes, including increasing, decreasing, bathtub-shaped, and modified bathtub-shaped forms. They also considered parameter estimation using maximum likelihood, least-squares, weighted least-squares and Bayesian approaches.

\vspace{0.25cm}
Although these results establish a solid foundation for the UT distribution, several important aspects remain unexplored. In particular, properties related to order statistics and L-moments have not yet been derived. Moreover, characterization results based on truncated moments, which play a key role in distinguishing probability models and understanding their structural behavior are currently unavailable. From an inferential viewpoint, a number of estimation techniques successfully applied to other unit interval distributions have not been examined for the UT distribution.

\vspace{0.25cm}
Motivated by these gaps, the present paper aims to further develop the theory and inference of the UT distribution. Explicit expressions for single moments of order statistics and L-moments are derived, and new characterization results based on truncated moments are established. In addition to the estimation methods previously studied, alternative procedures, including maximum product of spacings, ordinary and weighted least-squares, percentile based estimation, L-moments estimation, Anderson–Darling and right-tail Anderson–Darling estimation methods are investigated. The performance of these estimators is evaluated and compared through extensive Monte Carlo simulation studies under various parameter settings and sample sizes.

\vspace{0.25cm}
The remainder of the paper is organized as follows. Section~2 presents additional statistical properties of the UT distribution and two characterization results. Parameter estimation methods are discussed in Section~3, while Section~4 reports the results of the simulation study. A real data set is analyzed in Section~5. Concluding remarks are provided in Section~6.

\section{{Statistical Properties and Characterization Results}}

\subsection{Moments of order statistics}
Let $X_{1},\ldots,X_{n}$ be a random sample of size $n$ from the UT distribution with pdf $f(x;\theta)$ and cdf $F(x;\theta)$, given in \eqref{eq:1.1} and \eqref{eq:1.2}, respectively. Let $X_{1:n}\le \cdots \le X_{n:n}$ denote the corresponding order statistics. Then, for $1\le r\le n$, the pdf of the $r$th order statistic $X_{r:n}$ is given by (see David and Nagaraja \cite{DavidNagaraja2003}; Arnold et al. \cite{Arnold2008})
\begin{equation}\label{eq:2.1}
f_{r:n}(x;\theta)
=
\frac{n!}{(r-1)!(n-r)!}
[F(x;\theta)]^{r-1}
[1-F(x;\theta)]^{n-r}
f(x;\theta),
\quad 0<x<1.
\end{equation}

The $k$th moment of the $r$th order statistic is defined as
\begin{equation}\label{eq:2.2}
\mu_{r:n}^{(k)}
=
\mathbb{E}(X_{r:n}^{k})
=
\int_{0}^{1} x^{k} f_{r:n}(x;\theta)\,dx,
\quad 1\le r\le n,\; k\in\mathbb{N}.
\end{equation}

Explicit expressions for these moments are presented in the following theorems.

\subparagraph*{Theorem~2.1.}
For the UT distribution given in \eqref{eq:1.1} ($1 \le r \le n-1$, $k \in \mathbb{N}$), we have
\begin{equation}
\label{eq:2.3}
\mu^{(k)}_{r:n}
=
\sum_{i=r}^{n} (-1)^{i-r}
\binom{i-1}{r-1}
\binom{n}{i}
\frac{e^{i}}{i^{\,i-\frac{k}{\theta}}}
\left[
\Gamma\!\left(i-\frac{k}{\theta}+1, i\right)
- i\,\Gamma\!\left(i-\frac{k}{\theta}, i\right)
\right],
\end{equation}
where
\begin{equation}
\label{eq:2.4}
\Gamma(a,b) = \int_b^{\infty} t^{a-1} e^{-t} \, dt
\end{equation}
denotes the upper incomplete gamma function.\\

{\bf Proof.}~~In view of \eqref{eq:2.2} and the result given by David and Nagaraja \cite{DavidNagaraja2003} at page 45, we can write
\begin{equation}\label{eq:2.5}
\mu^{(k)}_{r:n}=\sum_{i=r}^{n}(-1)^{i-r}\binom{i-1}{r-1}\binom{n}{i}\mu^{(k)}_{i:i},~1\leq r\leq n-1,~k\in\mathbb{N},
\end{equation}
where
\begin{equation*}
\mu^{(k)}_{i:i}=i\int_{0}^{1}\!\!x^{k}{[F(x;\theta)]}^{i-1}{f(x;\theta)}{dx}
\end{equation*}
or, equivalently, from \eqref{eq:1.2} and \eqref{eq:1.3},
\begin{equation*}
\mu^{(k)}_{i:i}=i\theta\int_{0}^{1}\!\!{x^{k-1}}{(x^{-\theta}-1)}{[F(x;\theta)]}^{i}{dx}.
\end{equation*}
\vspace{-0.5cm}
\begin{equation}\label{eq:2.6}
\qquad=i\theta e^i\int_{0}^{1}\!\!{x^{k-i\theta-1}}e^{-ix^{-\theta}}{(x^{-\theta}-1)}{dx}
\end{equation}
Substituting $z=i x^{-\theta}$ into \eqref{eq:2.6} yields
\begin{equation*}
\mu^{(k)}_{i:i}=\frac{e^i}{i^{i-\frac{k}{\theta}}}\bigg[\int_{i}^{\infty}\!\!{z^{i-\frac{k}{\theta}}}e^{-z}{dz}
-i~\int_{i}^{\infty}\!\!{z^{i-\frac{k}{\theta}-1}}e^{-z}{dz}\bigg]
\end{equation*}
\begin{equation*}
\qquad=\frac{e^i}{i^{i-\frac{k}{\theta}}}\left[\Gamma\left(i-\frac{k}{\theta}+1,i\right)-i~\Gamma\left(i-\frac{k}{\theta},i\right)\right]
\end{equation*}
Inserting $\mu_{i:i}^{(k)}$ in \eqref{eq:2.5}, it follows \eqref{eq:2.3}.

\vspace{0.30 cm}
An alternative equation for $\mu_{r:n}^{(k)}$ is given in the following theorem.


\vspace{0.35cm}
{\bf Theorem~2.2.}~For $1\leq r\leq n$ and $k\in\mathbb{N}$, we obtain
\begin{equation*}
\mu^{(k)}_{r:n} = n\sum_{i =0}^{n-r}(-1)^i\binom{n\!-\!1}{r\!-\!1}\binom{n\!-\!r}{i}\frac{e^{i+r}}{(i+r)^{i+r-\frac{k}{\theta}+1}}
\end{equation*}
\vspace{-0.35cm}
\begin{equation}\label{eq:2.7}
\qquad\quad\times\left[\Gamma\left(i\!+\!r\!-\!\frac{k}{\theta}\!+\!1,i\!+\!r\right)
-(i\!+\!r)\Gamma\left(i\!+\!r\!-\!\frac{k}{\theta},i\!+\!r\right)\right],
\end{equation}
where $\Gamma(a,b)$ is defined as before.

\vspace{0.5cm}
{\bf Proof.}~~From \eqref{eq:2.2}, we have
\begin{eqnarray*}\label{(E3-1)}
\mu^{(k)}_{r:n}&=&C_{r:n}\int_{0}^{1}\!\! x^k [F(x;\theta)]^{r-1} [1-F(x;\theta)]^{n-r} f(x;\theta) {\rm d}x
\end{eqnarray*}
\begin{eqnarray}\label{(eq:2.8)}
\quad~&=&\label{pr1}C_{r:n}\sum_{i =0}^{n-r}(-1)^i~\binom{n-r}{i} \int_{0}^{1}\!\! x^k[F(x;\theta)]^{i+r} \frac{f(x;\theta)}{F(x;\theta)}\,{\rm d}x
\end{eqnarray}
Applying similar steps as in Theorem 2.1, leads to \eqref{eq:2.7}.

\vspace{0.35cm}
{\bf Remark 2.1.}~By setting $n=r=1$ in \eqref{eq:2.3} or \eqref{eq:2.7}, we obtain
\begin{equation}\label{eq:2.9}
\mu^{(k)}_{1:1}=E(X^k)=e\left[\Gamma\left(2-\frac{k}{\theta},1\right)-\Gamma\left(1-\frac{k}{\theta},1\right)\right],
\end{equation}
which corresponds to the $k$th moment of $X$ reported by Krishna et al. \cite{Krishna2022}.

\vspace{0.30cm}
By setting $k = 1, 2, 3, 4$ in \eqref{eq:2.9}, one may obtain closed form expressions for the first four moments of $X$.

\vspace{0.30cm}

\textcolor{black}{\bf Remark 2.2.} For $r=1$ in \eqref{eq:2.3}, we have
\begin{equation*}
\mu^{(k)}_{1:n} = n \sum_{i=0}^{n-1} (-1)^i \binom{n-1}{i} 
\frac{e^{i+1}}{(i+1)^{i-\frac{k}{\theta}+2}}
\Biggl[ 
\Gamma\left(i-\frac{k}{\theta}+2, i+1\right) - (i+1)\, \Gamma\left(i-\frac{k}{\theta}+1, i+1\right)
\Biggr],
\end{equation*}
and for $r=n$ in \eqref{eq:2.3}, we obtain
\begin{equation*}
\mu^{(k)}_{n:n} = \frac{n e^{n}}{n^{\,n-\frac{k}{\theta}+1}} 
\Bigl[ \Gamma\left(n-\frac{k}{\theta}+1, n\right) - n\, \Gamma\left(n-\frac{k}{\theta}, n\right) \Bigr],
\end{equation*}
which correspond to the $k$th moments of the minimum and maximum order statistics, respectively.

\subparagraph*{\textcolor{black}{Remark 2.3.} } By setting $k=1$ and $k=2$ in \eqref{eq:2.3}, the means and second moments of the order statistics for the UT distribution 
(for $n = 1, \dots, 5$) are computed for selected values of the shape parameter $\theta$, and are presented to six decimal places in Table~\ref{T:1}. 

\vspace{0.25cm}
The variance of $X_{r:n}$ ($1 \le r \le n$) is given by
\[
V(X_{r:n}) = \mu_{r:n}^{(2)} - \bigl[\mu_{r:n}^{(1)}\bigr]^2,
\]
where $\mu_{r:n}^{(1)}$ and $\mu_{r:n}^{(2)}$ denote the first and second moments, respectively. 

\vspace{0.25cm}
Table~\ref{T:1} presents the means, second moments and variances of the order statistics of the UT distribution for different sample sizes $n = 1(1)5$ and shape parameters $\theta = 1(1)4$. For a fixed sample size, both the means and second moments increase, while the variances decrease, as $\theta$ increases, indicating that larger shape parameters yield distributions more concentrated at higher values. For a fixed $\theta$, the means and second moments generally increase with the sample size, whereas the variances decrease, reflecting the stabilizing effect of larger samples on the dispersion of the order statistics.

\begin{table}[H]
\centering
\caption{Means, second moments and variances of order statistics for the UT distribution.}
\label{T:1}
\setlength{\tabcolsep}{1.2em}
\small
\begin{tabular}{ccccc|ccc}
\toprule
 & & \multicolumn{3}{c}{$\theta=1.0$} & \multicolumn{3}{c}{$\theta=2.0$} \\
\cmidrule(lr){3-5} \cmidrule(lr){6-8}
$n$ & $r$ & $\mu_{r:n}^{(1)}$ & $\mu_{r:n}^{(2)}$ & $V[X_{r:n}]$ & $\mu_{r:n}^{(1)}$ & $\mu_{r:n}^{(2)}$ & $V[X_{r:n}]$ \\
\midrule
1 & 1 & 0.40365 & 0.19269 & 0.02976 & 0.62106 & 0.40365 & 0.01793 \\
\midrule
2 & 1 & 0.30731 & 0.10805 & 0.01361 & 0.54480 & 0.30731 & 0.01050 \\
  & 2 & 0.50000 & 0.27734 & 0.02734 & 0.69733 & 0.50000 & 0.01373 \\
\midrule
3 & 1 & 0.26651 & 0.07939 & 0.00836 & 0.50895 & 0.26651 & 0.00748 \\
  & 2 & 0.38889 & 0.16536 & 0.01413 & 0.61649 & 0.38889 & 0.00883 \\
  & 3 & 0.55556 & 0.33333 & 0.02469 & 0.73775 & 0.55556 & 0.01128 \\
\midrule
4 & 1 & 0.24308 & 0.06506 & 0.00597 & 0.48701 & 0.24308 & 0.00590 \\
  & 2 & 0.33681 & 0.12239 & 0.00895 & 0.57478 & 0.33681 & 0.00644 \\
  & 3 & 0.44097 & 0.20833 & 0.01388 & 0.65820 & 0.44097 & 0.00774 \\
  & 4 & 0.59375 & 0.37500 & 0.02246 & 0.76426 & 0.59375 & 0.00965 \\
\midrule
5 & 1 & 0.22747 & 0.05638 & 0.00464 & 0.47173 & 0.22747 & 0.00494 \\
  & 2 & 0.30554 & 0.09976 & 0.00641 & 0.54813 & 0.30554 & 0.00509 \\
  & 3 & 0.38371 & 0.15633 & 0.00910 & 0.61474 & 0.38371 & 0.00580 \\
  & 4 & 0.47915 & 0.24300 & 0.01342 & 0.68718 & 0.47915 & 0.00693 \\
  & 5 & 0.62240 & 0.40800 & 0.02062 & 0.78353 & 0.62240 & 0.00848 \\
\midrule
 & & \multicolumn{3}{c}{$\theta=3.0$} & \multicolumn{3}{c}{$\theta=4.0$} \\
\cmidrule(lr){3-5} \cmidrule(lr){6-8}
$n$ & $r$ & $\mu_{r:n}^{(1)}$ & $\mu_{r:n}^{(2)}$ & $V[X_{r:n}]$ & $\mu_{r:n}^{(1)}$ & $\mu_{r:n}^{(2)}$ & $V[X_{r:n}]$ \\
\midrule
1 & 1 & 0.72416 & 0.53535 & 0.01094 & 0.78347 & 0.62106 & 0.00724 \\
\midrule
2 & 1 & 0.66445 & 0.44843 & 0.00693 & 0.73486 & 0.54480 & 0.00477 \\
  & 2 & 0.78387 & 0.62227 & 0.00782 & 0.83207 & 0.69733 & 0.00499 \\
\midrule
3 & 1 & 0.63543 & 0.40894 & 0.00518 & 0.71085 & 0.50895 & 0.00364 \\
  & 2 & 0.72249 & 0.52739 & 0.00540 & 0.78289 & 0.61649 & 0.00357 \\
  & 3 & 0.81456 & 0.66971 & 0.00620 & 0.85666 & 0.73775 & 0.00389 \\
\midrule
4 & 1 & 0.61730 & 0.38527 & 0.00421 & 0.69570 & 0.48701 & 0.00301 \\
  & 2 & 0.68981 & 0.47995 & 0.00412 & 0.75630 & 0.57478 & 0.00279 \\
  & 3 & 0.75517 & 0.57482 & 0.00454 & 0.80948 & 0.65820 & 0.00294 \\
  & 4 & 0.83436 & 0.70133 & 0.00519 & 0.87238 & 0.76426 & 0.00321 \\
\midrule
5 & 1 & 0.60450 & 0.36903 & 0.00361 & 0.68493 & 0.47173 & 0.00261 \\
  & 2 & 0.66851 & 0.45027 & 0.00336 & 0.73880 & 0.54813 & 0.00231 \\
  & 3 & 0.72175 & 0.52448 & 0.00355 & 0.78255 & 0.61474 & 0.00235 \\
  & 4 & 0.77745 & 0.60838 & 0.00396 & 0.82744 & 0.68718 & 0.00253 \\
  & 5 & 0.84858 & 0.72457 & 0.00448 & 0.88362 & 0.78353 & 0.00275 \\
\bottomrule
\end{tabular}
\end{table}
It can be seen that the condition $\sum_{r=1}^n \mu_{r:n}=nE(X)$ holds (see David and Nagaraja \cite{DavidNagaraja2003}).

\subsection{L-moments}
L-moments, introduced by Hosking \cite{Hosking1990}, are expectations of certain linear combinations of order statistics and provide robust alternatives to conventional moments. Let $X$ be a random variable with pdf given in \eqref{eq:1.1} and the corresponding cdf given in \eqref{eq:1.2}. The $m$th L-moment of $X$ is defined by
\begin{equation}\label{eq:2.10}
\lambda_m = \frac{1}{m} \sum_{j=0}^{m-1} (-1)^j \binom{m-1}{j} \, \mu_{m-j:m}, \quad m \ge 1,
\end{equation}
where
\begin{equation*}
\mu_{i:m} = \frac{m!}{(i-1)! (m-i)!} \int_0^1 x \, [F(x)]^{i-1} [1-F(x)]^{m-i} f(x) \, dx.
\end{equation*}  

The first four L-moments are obtained by setting $m=1,2,3,4$ in \eqref{eq:2.10}. For the UT distribution, they can be expressed explicitly as
\begin{align*}
\lambda_1 &= \mu_{1:1}, \\
\lambda_2 &= \mu_{2:2} - \mu_{1:1}, \\
\lambda_3 &= 2 \mu_{3:3} - 3 \mu_{2:2} + \mu_{1:1}, \\
\lambda_4 &= 5 \mu_{4:4} - 10 \mu_{3:3} + 6 \mu_{2:2} - \mu_{1:1},
\end{align*}
where
\begin{equation*}
\mu^{(k)}_{i:i} = \frac{e^i}{i^{i - k/\theta}} 
\left[ \Gamma\Big(i - \frac{k}{\theta} + 1, i \Big) - i \, \Gamma\Big(i - \frac{k}{\theta}, i \Big) \right].
\end{equation*}

L-moment ratios provide dimensionless measures of variability, asymmetry and kurtosis. In particular, the L-coefficient of variation (L-CV) is defined as
\[
\text{L-CV} = \frac{\lambda_2}{\lambda_1},
\]
while L-skewness and L-kurtosis are given by
\[
\tau_3 = \frac{\lambda_3}{\lambda_2}, \qquad
\tau_4 = \frac{\lambda_4}{\lambda_2}.
\]

The L-moments and associated ratios for the UT distribution were computed to six decimal places for selected parameter values and the results are reported in Table~\ref{T:2}.

\begin{table}[H]
\centering
\caption{First four L-moments and L-moment ratios for the UT distribution.}
\label{T:2}
\small
\begin{tabular}{ccccc}
\toprule
 & $\theta=1.0$ & $\theta=2.0$ & $\theta=3.0$ & $\theta=4.0$ \\
\midrule
$\lambda_1$ & 0.40365 & 0.62106 & 0.72416 & 0.78347 \\
$\lambda_2$ & 0.09635 & 0.07626 & 0.05971 & 0.04860 \\
$\lambda_3$ & 0.01476 & 0.00457 & 0.00167 & 0.00057 \\
$\lambda_4$ & 0.00954 & 0.00674 & 0.00524 & 0.00428 \\
\midrule
L-CV & 0.23869 & 0.12280 & 0.08245 & 0.06203 \\
L-skewness ($\tau_3$) & 0.15323 & 0.05997 & 0.02797 & 0.01183 \\
L-kurtosis ($\tau_4$) & 0.09904 & 0.08838 & 0.08781 & 0.08811 \\
\bottomrule
\end{tabular}
\end{table}

Table~\ref{T:2} presents the first four L-moments and the associated L-moment ratios of the UT distribution for selected values of $\theta$. It is evident that as $\theta$ increases, the mean ($\lambda_1$) increases, while the variability ($\lambda_2$), skewness ($\lambda_3$) and kurtosis ($\lambda_4$) decrease, indicating that the distribution becomes more concentrated around its mean and less asymmetric. This pattern is reinforced by the L-moment ratios, L-CV, L-skewness ($\tau_3$) and L-kurtosis ($\tau_4$), which also \textcolor{black}{decrease} with increasing $\theta$, confirming that higher values of $\theta$ yield a distribution that is both less dispersed and more symmetric. Overall, these results highlight the strong influence of $\theta$ on the shape and spread of the UT distribution.


\subsection{Characterizations}
Characterization results specify conditions under which a probability distribution is uniquely determined by its properties and play a central role in theoretical statistics, including model validation and the analysis of distributional properties. Approaches based on truncated moments and conditional expectations are particularly valuable due to their interpretability and analytical tractability (see Kagan et al. \cite{Kagan1973}, Galambos and Kotz \cite{Galambos1978}, Gl\"anzel \cite{Glanzel1987}).

\vspace{0.25cm}
We now present characterizations of the UT distribution based on truncated first moments. To prove these characterization results, we require two key lemmas and an assumption, which are stated below.

\vspace{0.35cm}
\noindent
\textbf{Assumption $\mathcal{A}$.}
Let $X$ be an absolutely continuous random variable with pdf given in \eqref{eq:1.1} and the corresponding cdf given in \eqref{eq:1.2}. Assume that $E(X)$ exists and that the density function $f(x)$ is differentiable. Let $\zeta = \inf\{x : F(x) > 0\}$ and $\eta = \sup\{x : F(x) < 1\}$
denote, respectively, the lower and upper endpoints of the support of $X$.

\vspace{0.35cm}
\noindent
\textbf{Lemma 3.1.}
Under Assumption $\mathcal{A}$, suppose that
\[
E(X \mid X \le x) = g(x)\,\tau(x),
\]
where $g(x)$ is continuous differentiable function of $x$ satisfying $\int_{\zeta}^{x} \frac{u - g'(u)}{g(u)}\,du$ is finite for all $x > \zeta$,
and $\tau(x) = \frac{f(x)}{F(x)}$. Then the pdf of $X$ is of the form
\[
f(x) = c \exp\!\left( \int \frac{x - g'(x)}{g(x)}\,dx \right),
\]
where the normalizing constant $c$ is uniquely determined by the condition
$\int_{\zeta}^{\eta} f(x)\,dx = 1$.

\vspace{0.35cm}
\noindent
\textbf{Lemma 3.2.}
Under Assumption $\mathcal{A}$, suppose that
\[
E(X \mid X \ge x) = h(x)\,r(x),
\]
where $h(x)$ is a continuous differentiable function of $x$ satisfying
$
\int_{\zeta}^{x} \frac{u - h'(u)}{h(u)}\,du$ is finite for all $x > \zeta$,
and $r(x) = \frac{f(x)}{1 - F(x)}$. Then the pdf of $X$ is of the form
\[
f(x) = c \exp\!\left( - \int \frac{x + h'(x)}{h(x)}\,dx \right),
\]
where the normalizing constant $c$ is uniquely determined by the condition
$\int_{\zeta}^{\eta} f(x)\,dx = 1$.

\vspace{0.25cm}
\noindent
The detailed proofs of Lemmas 3.1 and 3.2 are available in Ahsanullah \cite{Ahsanullah2017}.

\vspace{0.25cm}
We now present the characterization results for the UT distribution in the following theorems.

\vspace{0.35cm}
\noindent
\textbf{Theorem 3.1.}
Let $X$ be a random variable satisfying Assumption $\mathcal{A}$ with $\zeta = 0$ and $\eta = 1$. Then
\[
E(X \mid X \le x) = g(x)\,\tau(x)
\]
if and only if the pdf of $X$ is given by \eqref{eq:1.1}, where
$\tau(x) = \frac{f(x)}{F(x)}$ and
\begin{equation}\label{eq:2.11}
g(x)
=
\frac{
x^{1+2\theta}\,e^{x^{-\theta}}
}{
\theta\left(1 - x^{\theta}\right)
}
\left[
\Gamma\!\left(2 - \frac{1}{\theta}, \frac{1}{x^{\theta}}\right)
-
\Gamma\!\left(1 - \frac{1}{\theta}, \frac{1}{x^{\theta}}\right)
\right].
\end{equation}
Here, $\Gamma(\cdot,\cdot)$ is the upper incomplete gamma function as defined in \eqref{eq:2.4}.\\

\textbf{Proof.} 
Suppose $X$ has the pdf given by \eqref{eq:1.1}. Then
\[
g(x)\,\tau(x) = E(X \mid X \le x) = \frac{1}{F(x)} \int_0^x t\, f(t)\, dt.
\]
Since $\tau(x) = \frac{f(x)}{F(x)}$, it follows that
\begin{align*}
g(x)\,f(x) &= \int_0^x t\, \theta (t^{-\theta} - 1)\, t^{-(1+\theta)} e^{(1 - t^{-\theta})} \, dt \\ 
&= e \left\{
\Gamma\!\left(2 - \frac{1}{\theta}, \frac{1}{x^{\theta}}\right)
-
\Gamma\!\left(1 - \frac{1}{\theta}, \frac{1}{x^{\theta}}\right)
\right\}.
\end{align*}
Simplifying, we obtain
\[
g(x) = \frac{
x^{1 + 2\theta}\, e^{x^{-\theta}}
}{
\theta \,(1 - x^{\theta})
} \left\{
\Gamma\!\left(2 - \frac{1}{\theta}, \frac{1}{x^{\theta}}\right)
-
\Gamma\!\left(1 - \frac{1}{\theta}, \frac{1}{x^{\theta}}\right)
\right\}.
\]

Conversely, suppose $g(x)$ is given by \eqref{eq:2.11}. Differentiating $g(x)$ with respect to $x$ and simplifying, we obtain
\[
g'(x) = x - g(x) \left[ \frac{\theta}{x^{1+\theta}} - \frac{1 + 2\theta}{x} - \frac{\theta \, x^{\theta-1}}{1 - x^{\theta}} \right],
\]
so that
\[
\frac{x - g'(x)}{g(x)} = \frac{\theta}{x^{1+\theta}} - \frac{1 + 2\theta}{x} - \frac{\theta \, x^{\theta-1}}{1 - x^{\theta}}.
\]

By Lemma 3.1, it follows that
\[
\frac{f'(x)}{f(x)} = \frac{\theta}{x^{1+\theta}} - \frac{1 + 2\theta}{x} - \frac{\theta \, x^{\theta-1}}{1 - x^{\theta}}.
\]
Integrating both sides with respect to $x$ and using $\int_0^1 f(x)\, dx = 1$, we obtain
\[
f(x) = \theta \,(x^{-\theta} - 1) \, x^{-(1+\theta)} \, e^{1 - x^{-\theta}},~x \in (0,1), \ \theta > 0,
\]
which is the pdf given in \eqref{eq:1.1}. This completes the proof.\\

\textbf{Theorem 3.2.}
Suppose that the random variable $X$ satisfies Assumption $\mathcal{A}$ with $\zeta=0$ and $\eta=1$. Then
\[
E(X \mid X \ge x) = h(x)\,r(x)
\]
if and only if the pdf of X is given by (1.1), where $r(x)=\frac{f(x)}{1-F(x)}$ and
\begin{equation}\label{eq:2.13}
h(x)
=
\frac{
x^{1+2\theta}\,e^{x^{-\theta}}
}{
\theta\left(1 - x^{\theta}\right)
}
\left\{
\Gamma\!\left(2 - \frac{1}{\theta}, 1\right)
-
\Gamma\!\left(1 - \frac{1}{\theta}, 1\right)
-
\Gamma\!\left(2 - \frac{1}{\theta}, \frac{1}{x^{\theta}}\right)
+
\Gamma\!\left(1 - \frac{1}{\theta}, \frac{1}{x^{\theta}}\right)
\right\}.
\end{equation}

\vspace{0.35cm}
\textbf{Proof.} 
Suppose $X$ has the pdf given by \eqref{eq:1.1}. Then
\[
h(x)\,r(x) = E(X \mid X \ge x) = \frac{1}{1-F(x)} \int_x^1 t f(t)\, dt.
\]
Since $r(x) = \frac{f(x)}{1-F(x)}$, it follows that
\[
h(x) f(x) = E[X] - \int_0^x t f(t)\, dt
= e \left\{
\Gamma\!\left(2 - \frac{1}{\theta}, 1\right)
-
\Gamma\!\left(1 - \frac{1}{\theta}, 1\right)
-
\Gamma\!\left(2 - \frac{1}{\theta}, \frac{1}{x^{\theta}}\right)
+
\Gamma\!\left(1 - \frac{1}{\theta}, \frac{1}{x^{\theta}}\right)
\right\}.
\]
Hence, after simplification, we obtain (\ref{eq:2.13}).

Conversely, suppose $h(x)$ is given by (\ref{eq:2.13}). Differentiating $h(x)$ with respect to $x$ and simplifying, we obtain
\[
h'(x) = -x - h(x)\left[\frac{1+2\theta}{x}-\frac{\theta}{x^{1+\theta}}+\frac{\theta\,x^{\theta-1}}{1-x^{\theta}}\right].
\]
Hence,
\[
-\frac{x+h'(x)}{h(x)} = \frac{1+2\theta}{x}-\frac{\theta}{x^{1+\theta}}+\frac{\theta\,x^{\theta-1}}{1-x^{\theta}}.
\]
By Lemma 3.2, we have
\[
\frac{f'(x)}{f(x)} = \frac{1+2\theta}{x}-\frac{\theta}{x^{1+\theta}}+\frac{\theta\,x^{\theta-1}}{1-x^{\theta}}.
\]
Integrating both sides with respect to $x$ and using $\int_0^1 f(x)\, dx = 1$, we obtain
\[
f(x) = \theta(x^{-\theta}-1)x^{-(1+\theta)}e^{1-x^{-\theta}},~x \in (0,1), \ \theta>0.
\]
This completes the proof.

\section{Methods of Estimation}
In this section, we describe the different methods for estimating the parameter of the UT distribution. Using multiple estimation techniques is important, as it provides reliable parameter estimates and allows us to check how well the UT distribution fits different datasets.

\subsection{Maximum likelihood estimation}
\vspace{0.25cm}
Consider the random sample, say $X_1,X_2,\ldots,X_n$ of size $n$ from the UT distribution. Let $x_1, x_2, \dots, x_n$ denote the observed values of the sample. The log-likelihood function of  $\theta$ is given by
$$
\ell\left(\theta\right)=n\log{\left(\theta\right)}-\left(\theta+1\right)\sum_{i=1}^{n}{\log\ \left(x_i\right)}+\sum_{i=1}^{n}\log{\left({x_i}^{-\theta}-1\right)}-\sum_{i=1}^{n}\left({x_i}^{-\theta}-1\right).
$$

The maximum likelihood estimators (MLE) of $\theta$ can be obtained by simultaneously solving the following non-linear system:

$$
\frac{\partial\ell\left(\theta\right)}{\partial\theta}=\frac{n}{\theta}-\sum_{i=1}^{n}{\log\ \left(x_i\right)}-\sum_{i=1}^{n}{\frac{{x_i}^{-\theta}\log\ \left(x_i\right)}{{x_i}^{-\theta}-1}+\sum_{i=1}^{n}{{x_i}^{-\theta}\log\ \left(x_i\right)}}.
$$

\subsection{Ordinary least squares and weighted least squares estimation}
The least squares estimator (LSE) was proposed by Swain et al. \cite{Swain1988} to estimate the parameters of beta distributions. The LSE  of the parameter $\theta$  can be obtained by minimizing the following function:
\begin{eqnarray*}
L\left(\theta\right)=\sum_{i=1}^{n}\left[F\left(x_{i:n}\right)-\frac{i}{n+1}\right]^2
\end{eqnarray*}
\begin{eqnarray*}
\quad~&=&\sum_{i=1}^{n}{\left[{x_{i:n}}^{-\theta}e^{-{x_{i:n}}^{-\theta}+1}-\frac{i}{n+1}\right]^2.}
\end{eqnarray*}

An estimate of $\theta$  is calculated using the weighted LS estimator (WLSE) by minimizing the following function:
\begin{eqnarray*}
W\left(\theta\right)=\sum_{i=1}^{n}{\frac{\left(n+1\right)^2(n+2)}{i(n-i+1)}\left[F\left(x_{i:n}\right)-\frac{i}{n+1}\right]^2}
\end{eqnarray*}

\begin{eqnarray*}
\quad~&=&\sum_{i=1}^{n}{\frac{\left(n+1\right)^2(n+2)}{i(n-i+1)}\left[x_{i:n}^{-\theta}e^{-x_{i:n}^{-\theta}+1}-\frac{i}{n+1}\right]^2}.
\end{eqnarray*}

For more details, see Swain et al. \cite{Swain1988}.

\subsection{Maximum product of spacing estimation}
For estimating the unknown parameters of continuous univariate distributions, the maximum product of spacings (MPS) technique introduced by Cheng and Amin (\cite{ChengAmin1979}, \cite{ChengAmin1983}) presents a viable alternative to MLE. To estimate the parameters of the UT distribution using the MP estimators (MPSE), the following objective function must be minimized:
\begin{eqnarray*}
M\left(\theta\right)=\frac{1}{n+1}\sum_{i=1}^{n+1}\log{\left(\tau_i\right)},
\end{eqnarray*}
where
\begin{eqnarray*}
\tau_i=F\left(x_{i:n}\right)-F\left(x_{i-1:n}\right)=\left({x_{i:n}}^{-\theta}e^{-x_{i:n}^{-\theta}+1}\right)-\left({x_{i-1:n}}^{-\theta}e^{-x_{i-1:n}^{-\theta}+1}\right),
\end{eqnarray*}
\begin{eqnarray*}
F\left(x_{0:n}\right)=0\  \text{and}\ F\left(x_{n-1:n}\right)=1.
\end{eqnarray*}

\subsection{Cramér–von Mises estimation}
An important estimation approach discussed by Macdonald \cite{Macdonald1971} is the Cramér von--Mises estimation (CRVME) method. The CRVME estimator of the parameter $\theta$ can be obtained by minimizing the following function:
\begin{eqnarray*}
C\left(\theta\right)=\frac{1}{12n}+\sum_{i=1}^{n}\left[F\left(x_{i:n}\right)-\frac{2i-1}{2n}\right]^2
\end{eqnarray*}
\begin{eqnarray*}
\quad~&=&\frac{1}{12n}+\sum_{i=1}^{n}\left[x_{i:n}^{-\theta}e^{-x_{i:n}^{-\theta}+1}-\frac{2i-1}{2n}\right]^2.
\end{eqnarray*}

\subsection{Anderson–Darling and right-tail Anderson–Darling estimation}
The Anderson--Darling (AD) method is proposed by Anderson and Darling \cite{AndersonDarling1952}. The AD estimator (ADE) is another type of minimum distance estimator. The ADE of the parameters of the UT distribution are obtained by minimizing
\begin{eqnarray*}
A\left(\theta\right)=-n-\frac{1}{n}+\sum_{i=1}^{n}{\left(2i-1\right)\ \left[\log{F\left(x_{i:n}\right)}+\log{S\left(x_{i:n}\right)}\right]}
\end{eqnarray*}
\begin{eqnarray*}
\quad~&=&-n-\frac{1}{n}+\sum_{i=1}^{n}{\left(2i-1\right)\ \left[\log{(x_{i:n}^{-\theta}e^{-x_{i:n}^{-\theta}+1})}-\left(x_{i:n}^{-\theta}+1\right)+\theta\log{(x_{i:n})}\right].}
\end{eqnarray*}
On the other hand, Luce{\~n}o \cite{Luceno2006} applied some modifications on the AD statistic to define the right-tail Anderson–Darling estimators (RADE), which is specified by
\begin{eqnarray*}
R\left(\theta\right)=\frac{n}{2}-2\sum_{i=1}^{n}{\ F\left(x_{i:n}\right)-\frac{1}{n}\sum_{i=1}^{n}{\left(2i-1\right)\ \log{S\left(x_{i:n}\right)}}}
\end{eqnarray*}
\begin{eqnarray*}
\quad~&=&-\frac{n}{2}-2\sum_{i=1}^{n}{\ ({x_{i:n}}^{-\theta}e^{-x_{i:n}^{-\theta}+1})-\frac{1}{n}\sum_{i=1}^{n}\left(2i-1\right)\left[\theta\log{(x_{i:n})}-\left(x_{i:n}^{-\theta}+1\right)\right].}
\end{eqnarray*}

\subsection{Percentiles estimation}
Kao \cite{Kao1958} proposed the Percentile estimation (PCE) method. This method allows estimating the unknown parameters if the distribution function has a closed-form expression. \textcolor{black}{Let} $p=i/(n\ +\ 1)$ be an unbiased estimator of $F\left(x_{i:n}\right)$. \textcolor{black}{The} PCEs of the UT parameters are obtained by minimizing the following function
\begin{eqnarray*}
P\left(\theta\right)=\sum_{i=1}^{n}\left\{x_{i:n}-\left[-W_{-1}\left(\frac{p}
{e}\right)\right]^{-\frac{1}{\theta}}\right\}^2,
\end{eqnarray*}

where $p\in\left(0,1\right)$  and $W_{-1}$ denotes the negative branch of the Lambert W function.\\

\subsection{L-moment estimation}
Hosking \cite{Hosking1990} proposed an alternative estimation technique analogous to conventional moments, referred to as L-moment estimators (LME). These estimators are derived by equating the sample LME with their corresponding population counterparts. Hosking \cite{Hosking1990} highlighted that LME are generally more robust than traditional moment estimators, exhibiting greater resistance to the effects of outliers and maintaining reasonable efficiency compared to the MLE for certain distributions. The LME can be obtained by equating the first two sample LME with their corresponding population LME. The first two sample LME are defined as follows:
$$l_1 = \frac{1}{n} \sum_{i=1}^n l_{i:n}=\bar{x},\ \ \ l_2 = \frac{1}{2n(n-1)} \sum_{i=1}^n (i-1)l_{i:n}-l_{1},$$where $ l_{i:n} $ denotes the $i$th order statistic from a sample of size $ n $.
It follows $Q(p, \theta)$ that the first and second population LME are:
 \begin{equation*}
\mu_{1}(\theta)=\int_0^1 Q(p, \theta)dp=E(X|\theta)=e\left[\Gamma\left(2-\frac{k}{\theta},1\right)
-\Gamma\left(1-\frac{k}{\theta},1\right)\right]
\end{equation*}
 \begin{equation*}
\mu_{2}(\theta)=\int_0^1 Q(p, \theta) (2p-1)dp,
\end{equation*}
where $Q(p, \theta)$ quantile function.
The LME of $\theta$ is obtained as the simultaneous solution of the following system of equation:
$$
\mu_{1}(\hat{\theta})= l_1.
$$

\section{Simulation Study}

In this section, a Monte Carlo simulation study is conducted to evaluate the performance of various estimators of the unknown parameter of the UT distribution. The assessment is based on three widely used accuracy measures, namely the average absolute bias (BIAS), the mean squared error (MSE) and the average mean relative error (MRE), defined as
\begin{align*}
\text{BIAS}(\hat{\boldsymbol{\theta}}) &= \frac{1}{N}\sum_{i=1}^{N}
\left|\hat{\boldsymbol{\theta}}_{i}-\boldsymbol{\theta}\right|, \\
\text{MSE}(\hat{\boldsymbol{\theta}}) &= \frac{1}{N}\sum_{i=1}^{N}
\left(\hat{\boldsymbol{\theta}}_{i}-\boldsymbol{\theta}\right)^2, \\
\text{MRE}(\hat{\boldsymbol{\theta}}) &= \frac{1}{\boldsymbol{\theta}}
\left(\frac{1}{N}\sum_{i=1}^{N}
\left|\hat{\boldsymbol{\theta}}_{i}-\boldsymbol{\theta}\right|\right),
\end{align*}
respectively, where \(N\) denotes the number of Monte Carlo replications.

\vspace{0.25cm}
A total of \(N=1000\) random samples were generated from the UT distribution for sample sizes
\(n=\{30,50,100,250,500\}\) and parameter values
\(\theta=(0.26, 0.35, 0.5, 1, 1.25, 1.75, 2, 2.5, 3, 3.2)\).
For each combination of sample size and parameter value, the UT parameter was estimated using nine estimation methods: MLE, LSE, WLSE, MPSE, CRVME, ADE, RTADE, PCE and LME. Subsequently, the BIAS, MSE and MRE values were computed for each estimator. All simulations were performed using statistical software \textsf{R} (version~4.2.1) \cite{RCoreTeam2025}.

\vspace{0.25cm}
Tables~\ref{Table1}--\ref{Table10} present the results of the simulation study. In these tables, estimators are ranked row-wise, with individual ranks reported in braces \(\{\cdot\}\), while \(\sum \text{Ranks}\) denotes the cumulative rank of each estimator for a fixed sample size. The results in Tables~\ref{Table1}--\ref{Table10} indicate that all estimation methods are consistent, as the values of BIAS, MSE and MRE decrease with increasing sample size across all parameter configurations.

\vspace{0.25cm}
The partial and overall rankings of the estimators are summarized in Table~\ref{Table11}. The estimator with the smallest total rank is regarded as the best-performing method. Based on Table~\ref{Table11}, the nine estimation methods are ranked from best to worst as follows: MLE, MPSE, LME, ADE, WLSE, PCE, LSE, CRVME, and RTADE. Notably, the MLE method attains the lowest overall rank of 66.5, indicating its superior performance. These findings demonstrate that the MLE consistently outperforms the competing estimators in terms of BIAS, MSE and MRE across different sample sizes and parameter settings. Therefore, the simulation results indicate that the MLE method provides superior performance in estimating the parameters of the UT distribution.

\begin{table}[H]
\centering
\caption{Simulation results of the UT distribution for $\theta$=0.26.}
\label{Table1}
\resizebox{\linewidth}{!}{
\begin{tabular}{lllcccccccccc}
\toprule
 $n$         & Est.     & MLE      & LSE      & WLSE    & CRVME        & MPSE
             & PCE      & ADE      & RADE     & LME            \\ 
 \midrule
30  & $|BIAS|$ & 0.01464 $^{\{2\}}$   & 0.01813 $^{\{6\}}$   & 0.01691 $^{\{4\}}$   & 0.01831 $^{\{7\}}$   & 0.01528 $^{\{3\}}$   & 0.04836 $^{\{9\}}$ & 0.01736 $^{\{5\}}$   & 0.02023 $^{\{8\}}$ & 0.00408   $^{\{1\}}$   \\
    & MSE               & 0.00035 $^{\{1.5\}}$ & 0.00053 $^{\{5\}}$   & 0.00046 $^{\{3\}}$   & 0.00054 $^{\{6\}}$   & 0.00035 $^{\{1.5\}}$ & 0.00445 $^{\{9\}}$ & 0.00048 $^{\{4\}}$   & 0.00068 $^{\{7\}}$ & 0.00199   $^{\{8\}}$   \\
    & MRE               & 0.05630 $^{\{2\}}$   & 0.06974 $^{\{6\}}$   & 0.06503 $^{\{4\}}$   & 0.07041 $^{\{7\}}$   & 0.05875 $^{\{3\}}$   & 0.18599 $^{\{9\}}$ & 0.06675 $^{\{5\}}$   & 0.07779 $^{\{8\}}$ & 0.01570   $^{\{1\}}$   \\
    & $\sum{Ranks}$     & 5.5     $^{\{1\}}$   & 17      $^{\{6\}}$   & 11      $^{\{4\}}$   & 20      $^{\{7\}}$   & 7.5     $^{\{2\}}$   & 27      $^{\{9\}}$ & 14      $^{\{5\}}$   & 23      $^{\{8\}}$ & 10        $^{\{3\}}$
       \\
\midrule
50  & $|BIAS|$ & 0.01154 $^{\{2\}}$   & 0.01344 $^{\{6\}}$   & 0.01299 $^{\{5\}}$   & 0.01405 $^{\{7\}}$   & 0.01165 $^{\{3\}}$   & 0.03688 $^{\{9\}}$ & 0.01257 $^{\{4\}}$   & 0.01557 $^{\{8\}}$ & 0.00337   $^{\{1\}}$   \\
    & MSE               & 0.00022 $^{\{2\}}$   & 0.00030 $^{\{5\}}$   & 0.00027 $^{\{4\}}$   & 0.00032 $^{\{6\}}$   & 0.00020 $^{\{1\}}$   & 0.00225 $^{\{9\}}$ & 0.00025 $^{\{3\}}$   & 0.00040 $^{\{7\}}$ & 0.00128   $^{\{8\}}$   \\
    & MRE               & 0.04437 $^{\{2\}}$   & 0.05170 $^{\{6\}}$   & 0.04998 $^{\{5\}}$   & 0.05405 $^{\{7\}}$   & 0.04479 $^{\{3\}}$   & 0.14186 $^{\{9\}}$ & 0.04834 $^{\{4\}}$   & 0.05987 $^{\{8\}}$ & 0.00945   $^{\{1\}}$   \\
\bottomrule
\end{tabular}}
\end{table}

\begin{table}[H]
\centering
\caption{Simulation results of the UT distribution for $\theta$=0.35.}
\label{Table2}
\resizebox{\linewidth}{!}{
\begin{tabular}{lllcccccccccc}
\toprule
 $n$         & Est.     & MLE      & LSE      & WLSE    & CRVME        & MPSE
             & PCE      & ADE      & RADE     & LME            \\ 
\midrule
30  & $|BIAS|$ & 0.02043 $^{\{2\}}$   & 0.02334 $^{\{6\}}$   & 0.02324 $^{\{5\}}$ & 0.02457 $^{\{7\}}$   & 0.02060 $^{\{3\}}$   & 0.04936 $^{\{9\}}$ & 0.02320 $^{\{4\}}$   & 0.02632 $^{\{8\}}$ & 0.00668   $^{\{1\}}$ \\
    & MSE               & 0.00070 $^{\{3\}}$   & 0.00089 $^{\{5\}}$   & 0.00052 $^{\{1\}}$ & 0.00098 $^{\{6\}}$   & 0.00066 $^{\{2\}}$   & 0.00422 $^{\{9\}}$ & 0.00084 $^{\{4\}}$   & 0.00116 $^{\{7\}}$ & 0.00244   $^{\{8\}}$ \\
    & MRE               & 0.05839 $^{\{2\}}$   & 0.06670 $^{\{6\}}$   & 0.06641 $^{\{5\}}$ & 0.07021 $^{\{7\}}$   & 0.05887 $^{\{3\}}$   & 0.14103 $^{\{9\}}$ & 0.06628 $^{\{4\}}$   & 0.07521 $^{\{8\}}$ & 0.00491   $^{\{1\}}$ \\
    & $\sum{Ranks}$     & 7       $^{\{1\}}$   & 17      $^{\{6\}}$   & 11      $^{\{4\}}$ & 20      $^{\{7\}}$   & 8       $^{\{2\}}$   & 27      $^{\{9\}}$ & 12      $^{\{5\}}$   & 23      $^{\{8\}}$ & 10        $^{\{3\}}$
               \\
\midrule
50  & $|BIAS|$ & 0.01551 $^{\{2\}}$   & 0.01939 $^{\{7\}}$   & 0.01756 $^{\{5\}}$ & 0.01818 $^{\{6\}}$   & 0.01616 $^{\{3\}}$   & 0.03726 $^{\{9\}}$ & 0.01735 $^{\{4\}}$   & 0.02115 $^{\{8\}}$ & 0.00421   $^{\{1\}}$ \\
    & MSE               & 0.00039 $^{\{2\}}$   & 0.00057 $^{\{6\}}$   & 0.00024 $^{\{1\}}$ & 0.00053 $^{\{5\}}$   & 0.00040 $^{\{3\}}$   & 0.00216 $^{\{9\}}$ & 0.00049 $^{\{4\}}$   & 0.00072 $^{\{7\}}$ & 0.00117   $^{\{8\}}$ \\
    & MRE               & 0.04432 $^{\{2\}}$   & 0.05540 $^{\{7\}}$   & 0.05016 $^{\{5\}}$ & 0.05193 $^{\{6\}}$   & 0.04617 $^{\{3\}}$   & 0.10645 $^{\{9\}}$ & 0.04957 $^{\{4\}}$   & 0.06042 $^{\{8\}}$ & 0.00429   $^{\{1\}}$ \\
\bottomrule
\end{tabular}}
\end{table}

\begin{table}[H]
\centering
\caption{Simulation results of the UT distribution for $\theta$=0.5.}
\label{Table3}
\resizebox{\linewidth}{!}{
\begin{tabular}{lllcccccccccc}
\toprule
 $n$         & Est.     & MLE      & LSE      & WLSE    & CRVME        & MPSE
             & PCE      & ADE      & RADE     & LME            \\ 
\midrule
30  & $|BIAS|$ & 0.02846 $^{\{2\}}$   & 0.03440 $^{\{7\}}$ & 0.03307 $^{\{6\}}$   & 0.03228 $^{\{5\}}$ & 0.02946 $^{\{3\}}$   & 0.05320 $^{\{9\}}$ & 0.03143 $^{\{4\}}$   & 0.03730 $^{\{8\}}$ & 0.00353   $^{\{1\}}$ \\
    & MSE               & 0.00130 $^{\{1\}}$   & 0.00192 $^{\{6\}}$ & 0.00171 $^{\{4\}}$   & 0.00176 $^{\{5\}}$ & 0.00131 $^{\{2\}}$   & 0.00451 $^{\{9\}}$ & 0.00157 $^{\{3\}}$   & 0.00229 $^{\{7\}}$ & 0.00377   $^{\{8\}}$ \\
    & MRE               & 0.05692 $^{\{2\}}$   & 0.06880 $^{\{7\}}$ & 0.06613 $^{\{6\}}$   & 0.06456 $^{\{5\}}$ & 0.05892 $^{\{3\}}$   & 0.10641 $^{\{9\}}$ & 0.06285 $^{\{4\}}$   & 0.07461 $^{\{8\}}$ & 0.00705   $^{\{1\}}$ \\
    & $\sum{Ranks}$     & 5       $^{\{1\}}$   & 20      $^{\{7\}}$ & 16      $^{\{6\}}$   & 15      $^{\{5\}}$ & 8       $^{\{2\}}$   & 27      $^{\{9\}}$ & 11      $^{\{4\}}$   & 23      $^{\{8\}}$ & 10        $^{\{3\}}$
               \\
\midrule
50  & $|BIAS|$ & 0.02240 $^{\{3\}}$   & 0.02710 $^{\{6\}}$ & 0.02514 $^{\{5\}}$   & 0.02732 $^{\{7\}}$ & 0.02224 $^{\{2\}}$   & 0.04126 $^{\{9\}}$ & 0.02451 $^{\{4\}}$   & 0.03097 $^{\{8\}}$ & 0.00157   $^{\{1\}}$ \\
    & MSE               & 0.00080 $^{\{2\}}$   & 0.00120 $^{\{6\}}$ & 0.00100 $^{\{4\}}$   & 0.00114 $^{\{5\}}$ & 0.00074 $^{\{1\}}$   & 0.00273 $^{\{9\}}$ & 0.00095 $^{\{3\}}$   & 0.00151 $^{\{7\}}$ & 0.00194   $^{\{8\}}$ \\
    & MRE               & 0.04481 $^{\{3\}}$   & 0.05421 $^{\{6\}}$ & 0.05029 $^{\{5\}}$   & 0.05464 $^{\{7\}}$ & 0.04447 $^{\{2\}}$   & 0.08252 $^{\{9\}}$ & 0.04903 $^{\{4\}}$   & 0.06193 $^{\{8\}}$ & 0.00313   $^{\{1\}}$ \\
\bottomrule
\end{tabular}}
\end{table}

\begin{table}[H]
\centering
\caption{Simulation results of the UT distribution for $\theta$=1.}
\label{Table4}
\resizebox{\linewidth}{!}{
\begin{tabular}{lllcccccccccc}
\toprule
 $n$         & Est.     & MLE      & LSE      & WLSE    & CRVME        & MPSE
             & PCE      & ADE      & RADE     & LME            \\
 \midrule
30  & $|BIAS|$ & 0.06099 $^{\{3\}}$ & 0.07002 $^{\{7\}}$ & 0.06316 $^{\{5\}}$ & 0.06852 $^{\{6\}}$ & 0.06004 $^{\{2\}}$ & 0.07257 $^{\{8\}}$ & 0.06196 $^{\{4\}}$   & 0.07799 $^{\{9\}}$ & 0.01548   $^{\{1\}}$   \\
    & MSE               & 0.00586 $^{\{2\}}$ & 0.00807 $^{\{6\}}$ & 0.00647 $^{\{4\}}$ & 0.00773 $^{\{5\}}$ & 0.00540 $^{\{1\}}$ & 0.00839 $^{\{7\}}$ & 0.00625 $^{\{3\}}$   & 0.01030 $^{\{9\}}$ & 0.01006   $^{\{8\}}$   \\
    & MRE               & 0.06099 $^{\{3\}}$ & 0.07002 $^{\{7\}}$ & 0.06316 $^{\{5\}}$ & 0.06852 $^{\{6\}}$ & 0.06004 $^{\{2\}}$ & 0.07257 $^{\{8\}}$ & 0.06196 $^{\{4\}}$   & 0.07799 $^{\{9\}}$ & 0.01548   $^{\{1\}}$   \\
    & $\sum{Ranks}$     & 8       $^{\{2\}}$ & 20      $^{\{7\}}$ & 14      $^{\{5\}}$ & 17      $^{\{6\}}$ & 5       $^{\{1\}}$ & 23      $^{\{8\}}$ & 11      $^{\{4\}}$   & 27      $^{\{9\}}$ & 10        $^{\{3\}}$
               \\
\midrule
50  & $|BIAS|$ & 0.04326 $^{\{2\}}$ & 0.05193 $^{\{6\}}$ & 0.05053 $^{\{5\}}$ & 0.05259 $^{\{7\}}$ & 0.04427 $^{\{3\}}$ & 0.05701 $^{\{8\}}$ & 0.04919 $^{\{4\}}$   & 0.06077 $^{\{9\}}$ & 0.00954   $^{\{1\}}$   \\
    & MSE               & 0.00295 $^{\{1\}}$ & 0.00419 $^{\{5\}}$ & 0.00399 $^{\{4\}}$ & 0.00435 $^{\{6\}}$ & 0.00308 $^{\{2\}}$ & 0.00506 $^{\{7\}}$ & 0.00386 $^{\{3\}}$   & 0.00602 $^{\{9\}}$ & 0.00560   $^{\{8\}}$   \\
    & MRE               & 0.04326 $^{\{2\}}$ & 0.05193 $^{\{6\}}$ & 0.05053 $^{\{5\}}$ & 0.05259 $^{\{7\}}$ & 0.04427 $^{\{3\}}$ & 0.05701 $^{\{8\}}$ & 0.04919 $^{\{4\}}$   & 0.06077 $^{\{9\}}$ & 0.01270   $^{\{1\}}$   \\
\bottomrule
\end{tabular}}
\end{table}

\begin{table}[H]
\centering
\caption{Simulation results of the UT distribution for $\theta$=1.25.}
\label{Table5}
\resizebox{\linewidth}{!}{
\begin{tabular}{lllcccccccccc}
\toprule
 $n$         & Est.     & MLE      & LSE      & WLSE    & CRVME        & MPSE
             & PCE      & ADE      & RADE     & LME            \\ 
\midrule
30  & $|BIAS|$ & 0.07342 $^{\{3\}}$ & 0.08657 $^{\{7\}}$ & 0.08120 $^{\{5\}}$ & 0.08446 $^{\{6\}}$ & 0.07124 $^{\{2\}}$ & 0.08801 $^{\{8\}}$ & 0.07773 $^{\{4\}}$ & 0.09450 $^{\{9\}}$ & 0.00538   $^{\{1\}}$ \\
    & MSE               & 0.00891 $^{\{2\}}$ & 0.01249 $^{\{7\}}$ & 0.01068 $^{\{4\}}$ & 0.01200 $^{\{5\}}$ & 0.00806 $^{\{1\}}$ & 0.01238 $^{\{6\}}$ & 0.00980 $^{\{3\}}$ & 0.01516 $^{\{9\}}$ & 0.01397   $^{\{8\}}$ \\
    & MRE               & 0.05874 $^{\{3\}}$ & 0.06926 $^{\{7\}}$ & 0.06496 $^{\{5\}}$ & 0.06757 $^{\{6\}}$ & 0.05700 $^{\{2\}}$ & 0.07041 $^{\{8\}}$ & 0.06218 $^{\{4\}}$ & 0.07560 $^{\{9\}}$ & 0.00430   $^{\{1\}}$ \\
    & $\sum{Ranks}$     & 8       $^{\{2\}}$ & 21      $^{\{7\}}$ & 14      $^{\{5\}}$ & 17      $^{\{6\}}$ & 5       $^{\{1\}}$ & 22      $^{\{8\}}$ & 11      $^{\{4\}}$ & 27      $^{\{9\}}$ & 10        $^{\{3\}}$
               \\
\midrule
50  & $|BIAS|$ & 0.05600 $^{\{3\}}$ & 0.06636 $^{\{7\}}$ & 0.06409 $^{\{5\}}$ & 0.06871 $^{\{8\}}$ & 0.05502 $^{\{2\}}$ & 0.06550 $^{\{6\}}$ & 0.06149 $^{\{4\}}$ & 0.07342 $^{\{9\}}$ & 0.00798   $^{\{1\}}$ \\
    & MSE               & 0.00509 $^{\{2\}}$ & 0.00700 $^{\{6\}}$ & 0.00672 $^{\{4\}}$ & 0.00737 $^{\{7\}}$ & 0.00462 $^{\{1\}}$ & 0.00696 $^{\{5\}}$ & 0.00604 $^{\{3\}}$ & 0.00871 $^{\{9\}}$ & 0.00801   $^{\{8\}}$ \\
    & MRE               & 0.04480 $^{\{3\}}$ & 0.05309 $^{\{7\}}$ & 0.05127 $^{\{5\}}$ & 0.05497 $^{\{8\}}$ & 0.04402 $^{\{2\}}$ & 0.05240 $^{\{6\}}$ & 0.04919 $^{\{4\}}$ & 0.05874 $^{\{9\}}$ & 0.00406   $^{\{1\}}$ \\
\bottomrule
\end{tabular}}
\end{table}

\begin{table}[H]
\centering
\caption{Simulation results of the UT distribution for $\theta$=1.75.}
\label{Table6}
\resizebox{\linewidth}{!}{
\begin{tabular}{lllcccccccccc}
\toprule
 $n$         & Est.     & MLE      & LSE      & WLSE    & CRVME        & MPSE
             & PCE      & ADE      & RADE     & LME            \\ 
\midrule
30  & $|BIAS|$ & 0.09675 $^{\{2\}}$   & 0.12265 $^{\{8\}}$ & 0.11396 $^{\{5\}}$   & 0.11560 $^{\{7\}}$ & 0.10054 $^{\{3\}}$   & 0.11402 $^{\{6\}}$   & 0.10963 $^{\{4\}}$ & 0.13145 $^{\{9\}}$ & 0.01887   $^{\{1\}}$   \\
    & MSE               & 0.01506 $^{\{1\}}$   & 0.02587 $^{\{8\}}$ & 0.02151 $^{\{6\}}$   & 0.02172 $^{\{7\}}$ & 0.01587 $^{\{2\}}$   & 0.02025 $^{\{4\}}$   & 0.01943 $^{\{3\}}$ & 0.02947 $^{\{9\}}$ & 0.02073   $^{\{5\}}$   \\
    & MRE               & 0.05528 $^{\{2\}}$   & 0.07009 $^{\{8\}}$ & 0.06512 $^{\{5\}}$   & 0.06606 $^{\{7\}}$ & 0.05745 $^{\{3\}}$   & 0.06515 $^{\{6\}}$   & 0.06264 $^{\{4\}}$ & 0.07511 $^{\{9\}}$ & 0.00607   $^{\{1\}}$   \\
    & $\sum{Ranks}$     & 5       $^{\{1\}}$   & 24      $^{\{8\}}$ & 16      $^{\{5.5\}}$ & 21      $^{\{7\}}$ & 8       $^{\{3\}}$   & 16      $^{\{5.5\}}$ & 11      $^{\{4\}}$ & 27      $^{\{9\}}$ & 7         $^{\{2\}}$
               \\
\midrule
50  & $|BIAS|$ & 0.07672 $^{\{2\}}$   & 0.09124 $^{\{7\}}$ & 0.08469 $^{\{4\}}$   & 0.09762 $^{\{8\}}$ & 0.07837 $^{\{3\}}$   & 0.08893 $^{\{5\}}$   & 0.08922 $^{\{6\}}$ & 0.10495 $^{\{9\}}$ & 0.00955   $^{\{1\}}$   \\
    & MSE               & 0.00939 $^{\{1\}}$   & 0.01392 $^{\{7\}}$ & 0.01190 $^{\{3\}}$   & 0.01559 $^{\{8\}}$ & 0.00957 $^{\{2\}}$   & 0.01247 $^{\{4\}}$   & 0.01274 $^{\{5\}}$ & 0.01909 $^{\{9\}}$ & 0.01336   $^{\{6\}}$   \\
    & MRE               & 0.04384 $^{\{2\}}$   & 0.05214 $^{\{7\}}$ & 0.04839 $^{\{4\}}$   & 0.05578 $^{\{8\}}$ & 0.04478 $^{\{3\}}$   & 0.05082 $^{\{5\}}$   & 0.05098 $^{\{6\}}$ & 0.05997 $^{\{9\}}$ & 0.00545   $^{\{1\}}$   \\
\bottomrule
\end{tabular}}
\end{table}

\begin{table}[H]
\centering
\caption{Simulation results of the UT distribution for $\theta$=2.}
\label{Table7}
\resizebox{\linewidth}{!}{
\begin{tabular}{lllcccccccccc}
\toprule
 $n$         & Est.     & MLE      & LSE      & WLSE    & CRVME        & MPSE
             & PCE      & ADE      & RADE     & LME            \\ 
\midrule
30  & $|BIAS|$ & 0.11291 $^{\{2\}}$ & 0.13989 $^{\{8\}}$   & 0.13044 $^{\{6\}}$ & 0.13909 $^{\{7\}}$ & 0.11333 $^{\{3\}}$ & 0.12939 $^{\{5\}}$   & 0.12654 $^{\{4\}}$   & 0.15062 $^{\{9\}}$ & 0.02670   $^{\{1\}}$   \\
    & MSE               & 0.02079 $^{\{2\}}$ & 0.03183 $^{\{7\}}$   & 0.02823 $^{\{5\}}$ & 0.03162 $^{\{6\}}$ & 0.02010 $^{\{1\}}$ & 0.02630 $^{\{4\}}$   & 0.02612 $^{\{3\}}$   & 0.03834 $^{\{8\}}$ & 0.03963   $^{\{9\}}$   \\
    & MRE               & 0.05645 $^{\{2\}}$ & 0.06995 $^{\{8\}}$   & 0.06522 $^{\{6\}}$ & 0.06955 $^{\{7\}}$ & 0.05667 $^{\{3\}}$ & 0.06469 $^{\{5\}}$   & 0.06327 $^{\{4\}}$   & 0.07531 $^{\{9\}}$ & 0.01335   $^{\{1\}}$   \\
    & $\sum{Ranks}$     & 6       $^{\{1\}}$ & 23      $^{\{8\}}$   & 17      $^{\{6\}}$ & 20      $^{\{7\}}$ & 7       $^{\{2\}}$ & 14      $^{\{5\}}$   & 11      $^{\{3.5\}}$ & 26      $^{\{9\}}$ & 11        $^{\{3.5\}}$
               \\
\midrule
50  & $|BIAS|$ & 0.08905 $^{\{2\}}$ & 0.10432 $^{\{8\}}$   & 0.09798 $^{\{5\}}$ & 0.09943 $^{\{6\}}$ & 0.08966 $^{\{3\}}$ & 0.10266 $^{\{7\}}$   & 0.09513 $^{\{4\}}$   & 0.11843 $^{\{9\}}$ & 0.00980   $^{\{1\}}$   \\
    & MSE               & 0.01290 $^{\{2\}}$ & 0.01735 $^{\{7\}}$   & 0.01602 $^{\{4\}}$ & 0.01629 $^{\{5\}}$ & 0.01241 $^{\{1\}}$ & 0.01665 $^{\{6\}}$   & 0.01488 $^{\{3\}}$   & 0.02323 $^{\{9\}}$ & 0.02025   $^{\{8\}}$   \\
    & MRE               & 0.04452 $^{\{2\}}$ & 0.05216 $^{\{8\}}$   & 0.04899 $^{\{5\}}$ & 0.04971 $^{\{6\}}$ & 0.04483 $^{\{3\}}$ & 0.05133 $^{\{7\}}$   & 0.04757 $^{\{4\}}$   & 0.05921 $^{\{9\}}$ & 0.00490   $^{\{1\}}$   \\
\bottomrule
\end{tabular}}
\end{table}

\begin{table}[H]
\centering
\caption{Simulation results of the UT distribution for $\theta$=2.5.}
\label{Table8}
\resizebox{\linewidth}{!}{
\begin{tabular}{lllcccccccccc}
\toprule
 $n$         & Est.     & MLE      & LSE      & WLSE    & CRVME        & MPSE
             & PCE      & ADE      & RADE     & LME            \\ 
\midrule
30  & $|BIAS|$ & 0.13894 $^{\{2\}}$   & 0.17606 $^{\{8\}}$ & 0.17330 $^{\{7\}}$ & 0.16367 $^{\{6\}}$ & 0.14321 $^{\{3\}}$   & 0.15706 $^{\{5\}}$ & 0.15628 $^{\{4\}}$ & 0.19923 $^{\{9\}}$ & 0.01011   $^{\{1\}}$   \\
    & MSE               & 0.03173 $^{\{1.5\}}$ & 0.05048 $^{\{8\}}$ & 0.04682 $^{\{7\}}$ & 0.04556 $^{\{6\}}$ & 0.03173 $^{\{1.5\}}$ & 0.03949 $^{\{4\}}$ & 0.03943 $^{\{3\}}$ & 0.06824 $^{\{9\}}$ & 0.04195   $^{\{5\}}$   \\
    & MRE               & 0.05558 $^{\{2\}}$   & 0.07043 $^{\{8\}}$ & 0.06932 $^{\{7\}}$ & 0.06547 $^{\{6\}}$ & 0.05729 $^{\{3\}}$   & 0.06282 $^{\{5\}}$ & 0.06251 $^{\{4\}}$ & 0.07969 $^{\{9\}}$ & 0.00404   $^{\{1\}}$   \\
    & $\sum{Ranks}$     & 5.5     $^{\{1\}}$   & 24      $^{\{8\}}$ & 21      $^{\{7\}}$ & 18      $^{\{6\}}$ & 7.5     $^{\{3\}}$   & 14      $^{\{5\}}$ & 11      $^{\{4\}}$ & 27      $^{\{9\}}$ & 7         $^{\{2\}}$
               \\
\midrule
50  & $|BIAS|$ & 0.10806 $^{\{2\}}$   & 0.13153 $^{\{7\}}$ & 0.12144 $^{\{5\}}$ & 0.13555 $^{\{8\}}$ & 0.10944 $^{\{3\}}$   & 0.12408 $^{\{6\}}$ & 0.11627 $^{\{4\}}$ & 0.15029 $^{\{9\}}$ & 0.00109   $^{\{1\}}$   \\
    & MSE               & 0.01890 $^{\{2\}}$   & 0.02773 $^{\{6\}}$ & 0.02359 $^{\{4\}}$ & 0.02984 $^{\{8\}}$ & 0.01821 $^{\{1\}}$   & 0.02365 $^{\{5\}}$ & 0.02214 $^{\{3\}}$ & 0.03733 $^{\{9\}}$ & 0.02838   $^{\{7\}}$   \\
    & MRE               & 0.04323 $^{\{2\}}$   & 0.05261 $^{\{7\}}$ & 0.04858 $^{\{5\}}$ & 0.05422 $^{\{8\}}$ & 0.04378 $^{\{3\}}$   & 0.04963 $^{\{6\}}$ & 0.04651 $^{\{4\}}$ & 0.06011 $^{\{9\}}$ & 0.00244   $^{\{1\}}$   \\
\bottomrule
\end{tabular}}
\end{table}

\begin{table}[H]
\centering
\caption{Simulation results of the UT distribution for $\theta$=3.}
\label{Table9}
\resizebox{\linewidth}{!}{
\begin{tabular}{lllcccccccccc}
\toprule
 $n$         & Est.     & MLE      & LSE      & WLSE    & CRVME        & MPSE
             & PCE      & ADE      & RADE     & LME            \\ 
\midrule
30  & $|BIAS|$ & 0.17173 $^{\{2\}}$ & 0.21347 $^{\{8\}}$ & 0.19409 $^{\{6\}}$ & 0.20237 $^{\{7\}}$   & 0.17778 $^{\{3\}}$ & 0.18433 $^{\{4\}}$ & 0.18960 $^{\{5\}}$   & 0.23075 $^{\{9\}}$ & 0.02233   $^{\{1\}}$ \\
    & MSE               & 0.04754 $^{\{1\}}$ & 0.07751 $^{\{8\}}$ & 0.06154 $^{\{6\}}$ & 0.06603 $^{\{7\}}$   & 0.04888 $^{\{2\}}$ & 0.05445 $^{\{3\}}$ & 0.05772 $^{\{5\}}$   & 0.08818 $^{\{9\}}$ & 0.05718   $^{\{4\}}$ \\
    & MRE               & 0.05724 $^{\{2\}}$ & 0.07116 $^{\{8\}}$ & 0.06470 $^{\{6\}}$ & 0.06746 $^{\{7\}}$   & 0.05926 $^{\{3\}}$ & 0.06144 $^{\{4\}}$ & 0.06320 $^{\{5\}}$   & 0.07692 $^{\{9\}}$ & 0.00744   $^{\{1\}}$ \\
    & $\sum{Ranks}$     & 5       $^{\{1\}}$ & 24      $^{\{8\}}$ & 18      $^{\{6\}}$ & 21      $^{\{7\}}$   & 8       $^{\{3\}}$ & 11      $^{\{4\}}$ & 15      $^{\{5\}}$   & 27      $^{\{9\}}$ & 6         $^{\{2\}}$
               \\
\midrule
50  & $|BIAS|$ & 0.12714 $^{\{2\}}$ & 0.16228 $^{\{8\}}$ & 0.15343 $^{\{6\}}$ & 0.15527 $^{\{7\}}$   & 0.13212 $^{\{3\}}$ & 0.14653 $^{\{4\}}$ & 0.15088 $^{\{5\}}$   & 0.17953 $^{\{9\}}$ & 0.01453   $^{\{1\}}$ \\
    & MSE               & 0.02569 $^{\{1\}}$ & 0.04135 $^{\{8\}}$ & 0.03869 $^{\{6\}}$ & 0.03923 $^{\{7\}}$   & 0.02767 $^{\{2\}}$ & 0.03533 $^{\{3\}}$ & 0.03672 $^{\{4\}}$   & 0.05338 $^{\{9\}}$ & 0.03737   $^{\{5\}}$ \\
    & MRE               & 0.04238 $^{\{2\}}$ & 0.05409 $^{\{8\}}$ & 0.05114 $^{\{6\}}$ & 0.05176 $^{\{7\}}$   & 0.04404 $^{\{3\}}$ & 0.04884 $^{\{4\}}$ & 0.05029 $^{\{5\}}$   & 0.05984 $^{\{9\}}$ & 0.00484   $^{\{1\}}$ \\
\bottomrule
\end{tabular}}
\end{table}

\begin{table}[H]
\centering
\caption{Simulation results of the UT distribution for $\theta$=3.2.}
\label{Table10}
\resizebox{\linewidth}{!}{
\begin{tabular}{lllcccccccccc}
\toprule
 $n$         & Est.     & MLE      & LSE      & WLSE    & CRVME        & MPSE
             & PCE      & ADE      & RADE     & LME            \\ 
\midrule
30  & $|BIAS|$ & 0.18468 $^{\{3\}}$ & 0.21727 $^{\{7\}}$ & 0.21435 $^{\{6\}}$   & 0.22373 $^{\{8\}}$ & 0.18009 $^{\{2\}}$   & 0.19914 $^{\{4\}}$   & 0.20813 $^{\{5\}}$ & 0.25006 $^{\{9\}}$ & 0.04690   $^{\{1\}}$   \\
    & MSE               & 0.05442 $^{\{2\}}$ & 0.07367 $^{\{6\}}$ & 0.07365 $^{\{5\}}$   & 0.08048 $^{\{8\}}$ & 0.05172 $^{\{1\}}$   & 0.06269 $^{\{3\}}$   & 0.06928 $^{\{4\}}$ & 0.10191 $^{\{9\}}$ & 0.07995   $^{\{7\}}$   \\
    & MRE               & 0.05771 $^{\{3\}}$ & 0.06790 $^{\{7\}}$ & 0.06698 $^{\{6\}}$   & 0.06991 $^{\{8\}}$ & 0.05628 $^{\{2\}}$   & 0.06223 $^{\{4\}}$   & 0.06504 $^{\{5\}}$ & 0.07814 $^{\{9\}}$ & 0.01466   $^{\{1\}}$   \\
    & $\sum{Ranks}$     & 8       $^{\{2\}}$ & 20      $^{\{7\}}$ & 17      $^{\{6\}}$   & 24      $^{\{8\}}$ & 5       $^{\{1\}}$   & 11      $^{\{4\}}$   & 14      $^{\{5\}}$ & 27      $^{\{9\}}$ & 9         $^{\{3\}}$
               \\
\midrule
50  & $|BIAS|$ & 0.13492 $^{\{2\}}$ & 0.17390 $^{\{8\}}$ & 0.16053 $^{\{6\}}$   & 0.16744 $^{\{7\}}$ & 0.14579 $^{\{3\}}$   & 0.15866 $^{\{5\}}$   & 0.15478 $^{\{4\}}$ & 0.19270 $^{\{9\}}$ & 0.03717   $^{\{1\}}$   \\
    & MSE               & 0.03080 $^{\{1\}}$ & 0.05103 $^{\{7\}}$ & 0.04086 $^{\{5\}}$   & 0.04677 $^{\{6\}}$ & 0.03319 $^{\{2\}}$   & 0.04045 $^{\{4\}}$   & 0.03879 $^{\{3\}}$ & 0.05905 $^{\{9\}}$ & 0.05262   $^{\{8\}}$   \\
    & MRE               & 0.04216 $^{\{2\}}$ & 0.05434 $^{\{8\}}$ & 0.05017 $^{\{6\}}$   & 0.05233 $^{\{7\}}$ & 0.04556 $^{\{3\}}$   & 0.04958 $^{\{5\}}$   & 0.04837 $^{\{4\}}$ & 0.06022 $^{\{9\}}$ & 0.01162   $^{\{1\}}$   \\
\bottomrule
\end{tabular}}
\end{table}

\begin{table}[H]
\centering
\caption{Partial and overall rankings of the estimation methods for different values of $\theta$.}
\label{Table11}
\resizebox{0.80\linewidth}{!}{%
\begin{tabular}{lllcccccccccc}
\toprule
par.         & $n$      & MLE      & LSE      & WLSE    & CRVME        & MPSE
             & PCE      & ADE      & RADE     & LME            \\ 
\midrule
$\theta=0.26$ & 30  & 1    & 6   & 4     & 7     & 2    & 9     & 5   & 8    & 3     \\
              & 50  & 1    & 6   & 5     & 7     & 2    & 9     & 4   & 8    & 3     \\
              & 100 & 1    & 6   & 5     & 7     & 2    & 9     & 4   & 8    & 3     \\
              & 250 & 2    & 7   & 5     & 6     & 1    & 9     & 4   & 8    & 3     \\
              & 500 & 2    & 6   & 3.5   & 7     & 1    & 9     & 5   & 8    & 3.5   \\
\midrule
 $\theta=0.35$ & 30  & 1    & 6   & 4     & 7     & 2    & 9     & 5   & 8    & 3     \\
              & 50  & 1    & 7   & 4     & 6     & 2    & 9     & 5   & 8    & 3     \\
\bottomrule
\end{tabular}}
\end{table}


\section{Real Data Illustration}

In this section, we illustrate the importance of the UT distribution using a real-world dataset. The dataset is drawn from the cost-effectiveness dataset for corporate risk management and is defined as the ratio of total property, casualty, and uninsured premiums to total assets, yielding a bounded set within $(0,1)$. This dataset was previously analyzed by Gómez--Déniz et~al. \cite{GomezDeniz2014}. The observed data values are as follows:

\begin{table}[H]
\centering
\begin{tabular}{ccccc}
\hline
0.0278999962  & 0.0607999924 & 0.0215000095  & 0.0315000095 & 0.0040000006 \\
0.04070000172 & 0.1332999992 & 0.05820000172 & 0.0205999943 & 0.04309999943 \\
0.0694000057  & 0.0457999924 & 0.09130000114 & 0.0125       & 0.0036000014 \\
0.0534999905  & 0.1396000004 & 0.09          & 0.119200008  & 0.0782999924 \\
0.03289999962 & 0.1937999916 & 0.00930000007 & 0.1753000069 & 0.65 \\
0.0931000042  & 0.0375       & 0.10          & 0.14         & 0.08510000229 \\
0.2886000061  & 0.0028000001 & 0.1260999966  & 0.0064999976 & 0.05090000153 \\
0.04070000172 & 0.0411000134 & 0.1597000027  & 0.112899996  & 0.1356999969 \\
0.04340000153 & 0.0122000029 & 0.0432000172  & 0.0885000381 & 0.1244999981 \\
0.0432999924  & 0.0628999962 & 0.0848999771  & 0.007599999  & 0.0020000003 \\
0.15          & 0.0093999998 & 0.2221999931  & 0.0254999952 & 0.0370000048 \\
0.03890000105 & 0.0297000029 & 0.127100004   & 0.0818000305 & 0.15 \\
0.0528999962  & 0.0611999886 & 0.0525        & 0.0571000038 & 0.18 \\
0.9755000305  & 0.0216000086 & 0.2171999931  & 0.7930000305 & - \\
0.0790000095  & 0.0350999999 & 0.1832999992  & 0.2912000084 & - \\
\hline
\end{tabular}
\end{table}

Maximum likelihood estimates (MLEs) and their corresponding standard errors (SEs) are obtained for all model parameters. The goodness-of-fit of the competing models is evaluated using several standard information and performance criteria, including the negative maximized log-likelihood $-\hat{\ell}$, Akaike information criterion (AIC), consistent AIC (CAIC), Hannan--Quinn information criterion (HQIC), Bayesian information criterion (BIC), Cramér--von Mises statistic $(W^\ast)$, Anderson--Darling statistic $(A^\ast)$ and the Kolmogorov--Smirnov (KS) statistic with its associated $p$-value.

\vspace{0.25cm}
The best-fitting model is identified as the one exhibiting the largest KS $p$-value and the smallest values of IC, CAIC, BIC, HQIC, $W^\ast$, $A^\ast$, and KS statistics. For comparison, the following lifetime distributions are considered: Unit Burr--III (UBIII) (Modi and Gill \cite{ModiGill2020}), Unit-Gompertz (UG) (Mazucheli {et~al.} \cite{Mazucheli2019}), Kumaraswamy (Ku) (Kumaraswamy \cite{Kumaraswamy1980}), Beta (B) (Gupta and Nadarajah \cite{GuptaNadarajah2004}), Modified Kies Exponential (MKiE) (Al-Babtain et al. \cite{AlBabtain2020}), Log--Lindley (LL) (Gómez--Déniz et al. \cite{GomezDeniz2014}), Unit-Rayleigh (UR) (Bantan et al. \cite{Bantan2020}), New Power-Logarithmic (NPL) (Chesneau \cite{Chesneau2021}) and Unit Zeghdoudi (UZ) (Bashiru et al. \cite{Bashiru2025}).

\vspace{0.25cm}
The MLEs, SEs, and goodness-of-fit statistics for all models are summarized in Table~\ref{Table12}. The results indicate that the one-parameter UT distribution provides the best overall fit among the competing models.

\begin{table}[H]
\centering
\caption{Fitting measures, estimates and SEs for all models applied to the data}
\label{Table12}
\resizebox{\linewidth}{!}{%
\begin{tabular}{lcccccccccc}
\toprule
Model & MLEs (SEs) & $-\hat{\ell}$ & AIC & CAIC & BIC & HQIC & $W^*$ & $A^*$ & KS & $p$-value \\ 
\midrule
UL & $\hat{\theta}=0.3493\ (0.0155)$ & -88.5397 & -175.0790 & -175.0231 & -172.7889 & -174.1666 & 0.2220 & 1.4132 & 0.1033 & 0.4171 \\ 
UBIII & $\hat{\theta}=0.2334\ (0.0516)$ & -61.8314 & -119.6630 & -119.4915 & -115.0820 & -117.8374 & 0.5358 & 3.3881 & 0.3183 & 0.0000 \\ 
& $\hat{\beta}=1.5324\ (0.2966)$ & & & & & & & & & \\ 
UGo & $\hat{\theta}=0.1501\ (0.0546)$ & -87.1488 & -170.2977 & -170.1263 & -165.7168 & -168.4721 & 0.3149 & 1.8780 & 0.1311 & 0.1622 \\ 
& $\hat{\beta}=0.6047\ (0.0763)$ & & & & & & & & & \\ 
Ku & $\hat{\theta}=0.6648\ (0.0717)$ & -78.6539 & -153.3079 & -153.1364 & -148.7269 & -151.4823 & 0.4091 & 2.6575 & 0.1534 & 0.0641 \\ 
& $\hat{\beta}=3.4406\ (0.6208)$ & & & & & & & & & \\ 
B & $\hat{\theta}=0.6125\ (0.0855)$ & -76.1175 & -148.2350 & -148.0636 & -143.6541 & -146.4095 & 0.4789 & 3.0702 & 0.1805 & 0.0171 \\ 
& $\hat{\beta}=3.7978\ (0.7154)$ & & & & & & & & & \\ 
MKiEx & $\hat{\theta}=0.5790\ (0.0547)$ & -73.2304 & -142.4608 & -142.2894 & -137.8799 & -140.6352 & 0.4466 & 2.9199 & 0.2506 & 0.0002 \\ 
& $\hat{\beta}=3.9215\ (0.4396)$ & & & & & & & & & \\ 
LL & $\hat{\theta}=0.0342\ (0.0611)$ & -76.6041 & -149.2080 & -149.0369 & -144.6274 & -147.3827 & 0.2970 & 1.9823 & 0.2259 & 0.0011 \\ 
& $\hat{\beta}=0.6906\ (0.0588)$ & & & & & & & & & \\ 
UR & $\hat{\theta}=0.1034\ (0.0121)$ & -86.6386 & -171.2774 & -171.2210 & -168.9869 & -170.3646 & 0.2124 & 1.4162 & 0.1479 & 0.0819 \\ 
NPL & $\hat{\theta}=0.6592\ (0.0899)$ & -45.3172 & -88.6345 & -88.5781 & -86.3440 & -87.7217 & 0.5333 & 3.4059 & 0.2971 & 0.0000 \\ 
UZ & $\hat{\theta}=3.3326\ (0.2414)$ & 191.4548 & 384.9096 & 384.9660 & 387.2001 & 385.8224 & 1.1564 & 6.6561 & 0.7674 & 0.0000 \\ 
\bottomrule
\end{tabular}}
\end{table}

\begin{figure}[H]
	\centering
	\includegraphics[width=1\textwidth, height=0.4\textheight]{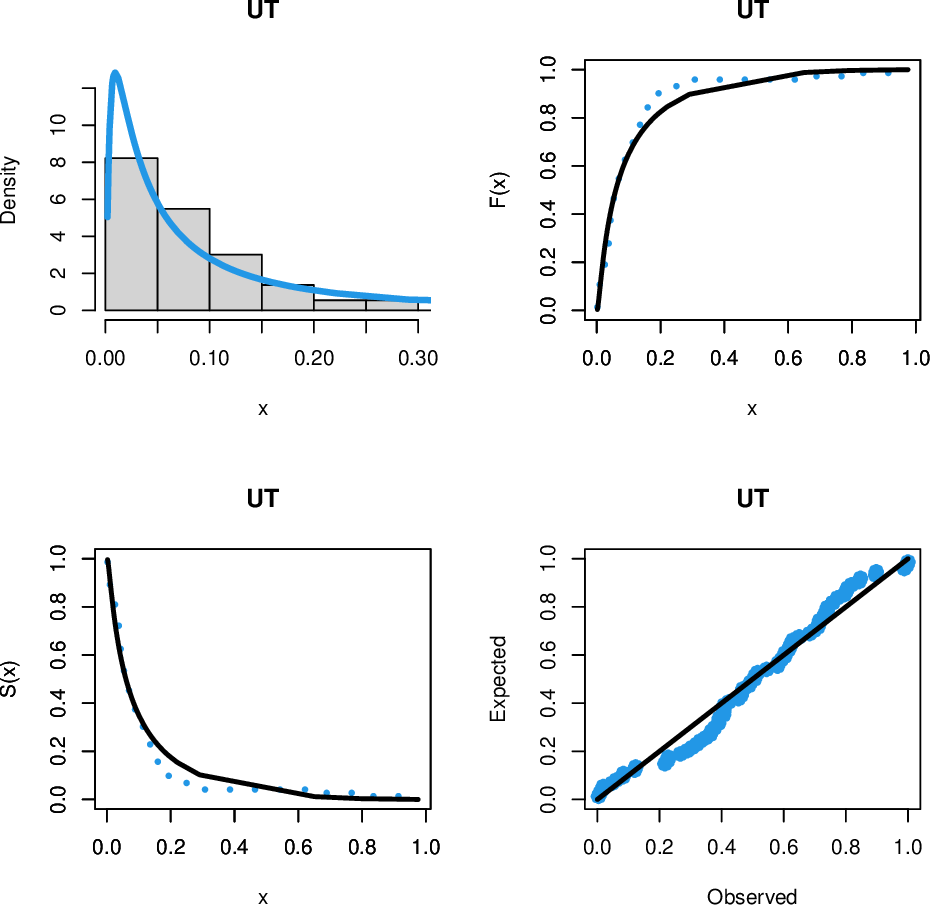}
	\hspace*{0.1cm}
	\caption{\footnotesize{Fitted pdf, cdf, sf and pp plots of the UT distribution for the data set.}}
\label{fig5}
\end{figure}

\begin{figure}[h!]
	\centering
	\includegraphics[width=1\textwidth, height=0.4\textheight]{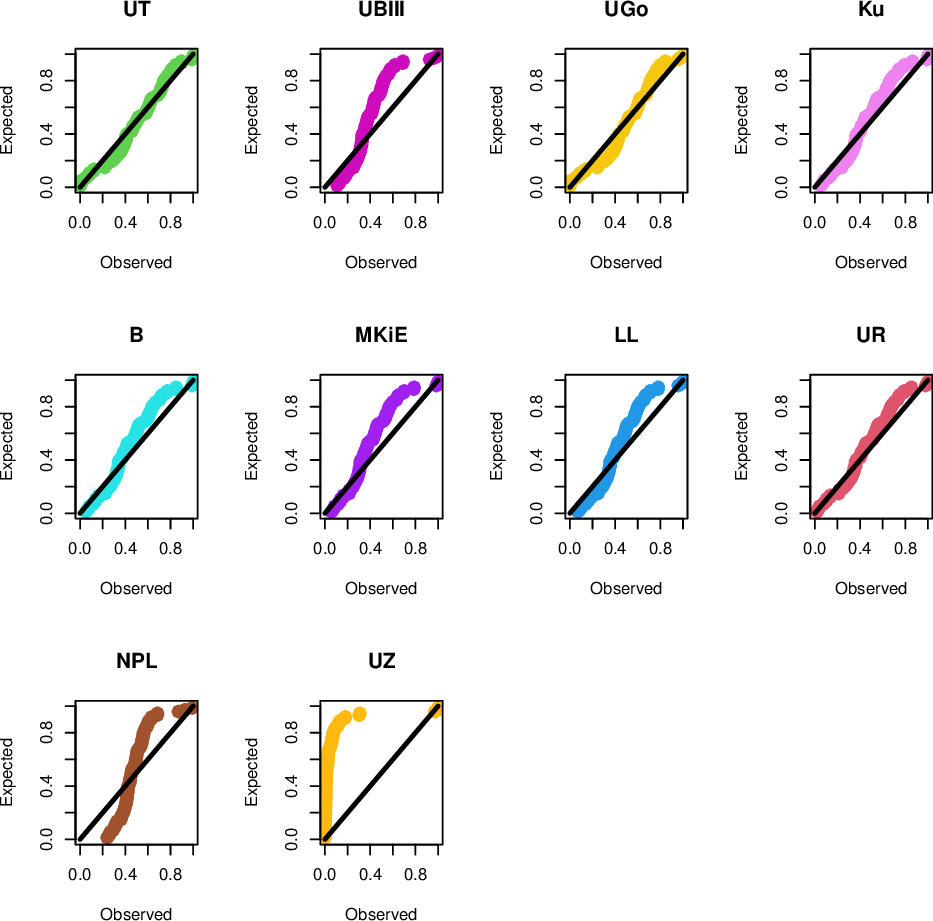}
	\hspace*{0.1cm}
\caption{\footnotesize{pp plots comparing the UT distribution with competing distributions for the dataset.}}

\label{fig6}
\end{figure}

Figure~\ref{fig5} presents the fitted pdf, cdf, survival function (sf) and probability--probability (pp) plots of the UT distribution. In addition, the pp plots shown in Figure~\ref{fig6} further support the superior fit of the UT distribution relative to the alternative models. These graphical results are consistent with the numerical findings reported in Table~\ref{Table12}.

\section*{Conclusions}

In this paper, further theoretical and inferential results for the Unit Teissier distribution \textcolor{black}{are} developed. Explicit expressions for single moments of order statistics and L-moments \textcolor{black}{are} obtained, and characterization results based on truncated moments \textcolor{black}{are}  established. These results add to the existing literature and provide additional insight into the structure of the UT distribution.

\vspace{0.25cm}
Several methods for parameter estimation \textcolor{black}{are} examined through simulation. In addition to maximum likelihood, least squares, weighted least squares, and Bayesian methods studied earlier, this paper considered maximum product of spacings, Cramér--von Mises, Anderson--Darling, right-tail Anderson--Darling, percentile, and L-moment estimators. The simulation results show that the maximum likelihood estimator performs best overall in terms of bias and error measures for the settings considered. An application to a corporate risk management dataset illustrates that the UT distribution can provide an adequate fit to bounded data encountered in practice.

\vspace{0.25cm}
There are several possible directions for future work. A location--scale family based on the UT distribution could be developed to increase modeling flexibility. The results obtained for order statistics may also be used to develop linear inference procedures. Other extensions, such as regression models, multivariate versions, and robust inference methods may be explored in future studies.\\
\\
\\

{\bf Declarations}\\

{\bf CRediT authorship contribution statement:}\\

{\bf Zuber Akhter:} Conceptualization of this study, Methodology, Software, Writing.\\
{\bf Mohamed A. Abdelaziz:} Conceptualization of this study, Methodology, Software, Writing.\\
{\bf M.Z. Anis:} Conceptualization of this study, Methodology, Writing.\\
{\bf Ahmed Z. Afify:} Conceptualization of this study, Methodology, Software, Writing.\\

{\bf Conflict of interest:} The authors declare that they have no conflict of interest.\\

{\bf Ethical statement:} The authors confirm their adherence to publication ethics and declare that this work is original and has not been previously published.\\

{\bf Funding statement:} Not applicable.


\begin{thebibliography}{99}

\bibitem{Afuecheta2025}
Afuecheta, E., Okorie, I.E., Jallow, H. and Nadarajah, S. (2025).
A review of unit continuous probability distributions.
{\em AIMS Mathematics}, {\bf 10}(11), 25939--26057.

\bibitem{Ahsanullah2017}
Ahsanullah, M. (2017).
{\em Characterizations of Univariate Continuous Distributions}.
Atlantis Press, Paris.

\bibitem{AlBabtain2020}
Al-Babtain, A.A., Shakhatreh, M.K., Nassar, M. and Afify, A.Z. (2020).
A new modified Kies family: Properties, estimation under complete and type-II censored samples and engineering applications.
{\em Mathematics}, {\bf 8}(8), 1345.

\bibitem{AndersonDarling1952}
Anderson, T.W. and Darling, D.A. (1952).
Asymptotic theory of certain ``goodness of fit'' criteria based on stochastic processes.
{\em Ann. Math. Statist.}, {\bf 23}(2), 193--212.

\bibitem{Arnold2008}
Arnold, B.C., Balakrishnan, N. and Nagaraja, H.N. (2008).
{\em A First Course in Order Statistics}.
SIAM, Philadelphia.

\bibitem{Bantan2020}
Bantan, R. A., Chesneau, C., Jamal, F., Elgarhy, M., Tahir, M.H., Ali, A., Zubair, M. and Anam, S. (2020).
Some new facts about the unit-Rayleigh distribution with applications.
{\em Mathematics}, {\bf 8}(11), 1954.

\bibitem{Bashiru2025}
Bashiru, S.O., Kayid, M., Sayed, R.M., Balogun, O.S., Abd El-Raouf, M.M. and Gemeay, A.M. (2025).
Introducing the unit Zeghdoudi distribution as a novel statistical model for analyzing proportional data.
{\em J. Radiat. Res. Appl. Sci.}, {\bf 18}(1), 101204.

\bibitem{ChengAmin1979}
Cheng, R.C.H. and Amin, N.A.K. (1979).
Maximum product of spacings estimation with applications to the log-Normal distribution.
Math Report 79--1, University of Wales Institute of Science and Technology, Cardiff.

\bibitem{ChengAmin1983}
Cheng, R.C.H., and Amin, N.A.K. (1983).
Estimating parameters in continuous univariate distributions with a shifted origin.
{\em J. R. Stat. Soc. Ser. B. Stat. Methodol.}, {\bf 45}(3), 394--403.

\bibitem{Chesneau2021}
Chesneau, C. (2021).
Study of a unit power-logarithmic distribution.
{\em Open J. Math. Sci.}, {\bf 5}, 218--235.

\bibitem{DavidNagaraja2003}
David, H.A. and Nagaraja, H.N. (2003).
{\em Order Statistics}, 3rd ed.
John Wiley, New York.

\bibitem{Galambos1978}
Galambos, J. and Kotz, S. (1978).
Characterizations of Probability Distributions: A Unified Approach with an Emphasis on Exponential and Related Models.
{\em Lecture Notes in Mathematics}, Vol. 675, Springer, Berlin.

\bibitem{GomezDeniz2014}
Gómez-Déniz, E., Sordo, M.A. and Calderín-Ojeda, E. (2014).
The Log--Lindley distribution as an alternative to the beta regression model with applications in insurance.
{\em Insurance Math. Econom.}, {\bf 54}, 49--57.

\bibitem{GuptaNadarajah2004}
Gupta, A.K. and Nadarajah, S. (2004).
{\em Handbook of Beta Distribution and Its Applications}.
Marcel Dekker, New York.

\bibitem{Hosking1990}
Hosking, J.R.M. (1990).
L-moments: analysis and estimation of distributions using linear combinations of order statistics.
{\em J. R. Stat. Soc. Ser. B. Stat. Methodol.}, {\bf 52}(1), 105--124.

\bibitem{Johnson1995}
Johnson, N.L., Kotz, S. and Balakrishnan, N. (1995).
{\em Continuous Univariate Distributions}, Vol. 2.
John Wiley \& Sons, New York.

\bibitem{Kagan1973}
Kagan, A.M., Linnik, Yu.V. and Rao, C.R. (1973).
{\em Characterization Problems in Mathematical Statistics}.
John Wiley \& Sons, New York.

\bibitem{Kao1958}
Kao, J. H. (1958).
Computer methods for estimating Weibull parameters in reliability studies.
{\em IRE Transactions on Reliability and Quality Control}, {\bf 13}, 15--22.

\bibitem{Krishna2022}
Krishna, A., Maya, R., Chesneau, C. and Irshad, M.R. (2022).
The Unit Teissier distribution and its applications.
{\em Math. Comput. Appl.}, {\bf 27}(1), 12.

\bibitem{Kumaraswamy1980}
Kumaraswamy, P. (1980).
A generalized probability density function for double-bounded random processes.
{\em Journal of Hydrology}, {\bf 46}(1--2), 79--88.

\bibitem{Luceno2006}
Luceño, A. (2006).
Fitting the generalized Pareto distribution to data using maximum goodness-of-fit estimators.
{\em Comput. Statist. Data Anal.}, {\bf 51}(2), 904--917.

\bibitem{Macdonald1971}
Macdonald, P.D.M. (1971).
Comment on ``An estimation procedure for mixtures of distributions'' by Choi and Bulgren.
{\em J. R. Stat. Soc. Ser. B. Stat. Methodol.}, {\bf 33}(2), 326--329.

\bibitem{Mazucheli2019}
Mazucheli, J., Menezes, A.F. and Dey, S. (2019).
Unit-Gompertz distribution with applications.
{\em Statistica}, {\bf 79}(1), 25--43.

\bibitem{ModiGill2020}
Modi, K. and Gill, V. (2020).
Unit Burr-III distribution with application.
{\em Journal of Statistics and Management Systems}, {\bf 23}(3), 579--592.

\bibitem{Nadarajah2007}
Nadarajah, S. and Kotz, S. (2007).
Multitude of beta distributions with applications.
{\em Statistics}, {\bf 41}(2), 153--179.

\bibitem{Nakamura2019}
Nakamura, L.R., Cerqueira, P.H.R., Ramires, T.G., Pescim, R.R., Rigby, R.A. and Stasinopoulos, D.M. (2019).
A new continuous distribution on the unit interval applied to modelling the points ratio of football teams.
{\em J. Appl. Stat.}, {\bf 46}(3), 416--431.

\bibitem{Ormoz2026}
Ormoz, E., Akhter, Z., Alam, M. and MirMostafaee, S.M.T.K. (2026).
Lower k-record values from unit-Gompertz distribution and associated inference.
{\em J. Stat. Comput. Simul.}, {\bf 96}(3), 532--562.

\bibitem{RCoreTeam2025}
R Core Team (2025).
{\em R: A Language and Environment for Statistical Computing}.
R Foundation for Statistical Computing, Vienna.

\bibitem{Sarhan2025}
Sarhan, A.M. (2025).
Unit-Chen distribution and its quantile regression model with applications.
{\em Scientific African}, {\bf 27}, e02555.

\bibitem{Swain1988}
Swain, J.J., Venkatraman, S. and Wilson, J.R. (1988).
Least squares estimation of distribution function in Johnson's translation system.
{\em J. Stat. Comput. Simul.}, {\bf 29}(4), 271--297.

\bibitem{Teissier1934}
Teissier, G. (1934).
Recherches sur le vieillissement et sur les lois de la mortalité.
{\em Ann. Physiol. Physicochim. Biol.}, {\bf 10}, 237--284.

\bibitem{Glanzel1987}
Gl\"anzel, W. (1987).
A characterization theorem based on truncated moments and its application to some distribution families.
In: Bauer, P., Konečný, F. and Wertz, W. (eds.),
{\em Mathematical Statistics and Probability Theory}, Vol. B, pp. 75--84.
Reidel, Dordrecht.

\end{thebibliography}
\end{document}